\documentclass[twocolumn,nofootinbib,superscriptaddress]{revtex4}
\usepackage{float}
\usepackage{graphicx}% Include figure files
\usepackage{dcolumn}% Align Tab.~columns on decimal point
\usepackage{bm}% bold math
\usepackage{amsmath}
\usepackage{hyperref}
\usepackage{xcolor}
\usepackage{subfigure}
\usepackage{verbatim}
\usepackage{amssymb}
\usepackage{ulem}

\renewcommand{\d}{\mathrm{d}}

\newcommand{\true}{\mathrm{true}}
\newcommand{\obs}{\mathrm{obs}}
\newcommand{\mock}{\mathrm{mock}}
\newcommand{\LR}{\mathrm{LR}}
\newcommand{\reio}{\mathrm{reio}}
\newcommand{\cdm}{\mathrm{cdm}}

\newcommand{\data}{\vec x}
\newcommand{\allpar}{\vec{\theta}}
\newcommand{\poi}{\vec \mu}
\newcommand{\onepoi}{\mu}
\newcommand{\np}{\vec \nu}

\newcommand{\Planck}{\textit{Planck}}

\newcommand{\pinc}{\texttt{pinc}}
\newcommand{\CLASS}{\texttt{CLASS}}
\newcommand{\Pliklite}{\texttt{Plik\_lite}}
\newcommand{\Montepython}{\texttt{MontePython}}

\renewcommand{\emph}{\textit}
\newcommand{\hhat}[1]{\hat{\hat{#1}}}
\renewcommand{\vec}[1]{\boldsymbol{#1}}
\newcommand{\like}{\mathcal{L}}
\newcommand{\LCDM}{$\Lambda$CDM}

\begin{document}

\title{Profile Likelihoods in Cosmology: \\
When, Why and How illustrated with $\Lambda$CDM, Massive Neutrinos and Dark Energy}

\author{Laura Herold}
\email{lherold@jhu.edu}
\affiliation{Department of Physics and Astronomy, Johns Hopkins University,
3400 North Charles Street, Baltimore, Maryland 21218, USA}

\author{Elisa G. M. Ferreira}
\affiliation{Kavli Institute for the Physics and Mathematics of the Universe (WPI), UTIAS, The University of Tokyo, Chiba 277-8583, Japan}

\author{Lukas Heinrich}
\affiliation{Technical University of Munich, TUM School of Natural Sciences, Physics Department, 85747 Garching, Germany}

\begin{abstract}\noindent
    Frequentist parameter inference using profile likelihoods has received increased attention in the cosmology literature recently since it can give important complementary information to Bayesian credible intervals. 
    Here, we give a pedagogical review of frequentist parameter inference in cosmology and focus on when the graphical profile likelihood construction gives meaningful constraints, i.e.\ confidence intervals with correct coverage.
    This construction rests on the assumption of the asymptotic limit of a large data set such as in \textit{Wilks' theorem}. 
    We assess the validity of this assumption in the context of three cosmological models with \textit{Planck} 2018 \texttt{Plik\_lite} data: While our tests for the $\Lambda$CDM model indicate that the profile likelihood method gives correct coverage, $\Lambda$CDM with the sum of neutrino masses as a free parameter appears consistent with a Gaussian near a boundary motivating the use of the boundary-corrected or Feldman-Cousins graphical method; for $w_0$CDM with the equation of state of dark energy, $w_0$, as a free parameter, we find indication of a violation of the assumptions. Finally, we compare frequentist and Bayesian constraints of these models. 
    Our results motivate care when using the graphical profile likelihood method in cosmology. Along with this paper, we publish our profile-likelihood code \href{https://github.com/LauraHerold/pinc}{\texttt{pinc}}.
\end{abstract}

\maketitle

%%%%%%%%%%%%%%%%%%%%%%%%%%%%%%%%%%%%%%%%%%%%%%%%%%%%%%%%%%%%%%%%%%%%%%

\section{Introduction}
\label{sec:intro}

Cosmology as a precision science owes much of its progress to the large and precise cosmological data sets, which led to the standard cosmological model, the $\Lambda$ cold dark matter ($\Lambda$CDM) model~\cite{Komatsu:2014ioa, Planck:2018vyg}. At the core of this achievement is the statistical analysis of these data sets, with Bayesian statistics playing a key role, facilitated by Markov chain Monte Carlo (MCMC) techniques~\cite{Christensen:2001gj} which allows to efficiently handle the high-dimensional parameter spaces common in cosmology. 

Future observations will provide larger, more precise data sets, finally shedding light on the dark components of the Universe~\cite{Verde:2019ivm,Abdalla:2022yfr}. However, the increased statistical power and complexity from new observational, theoretical, and systematic uncertainties introduce additional nuisance parameters, making statistical inference more challenging. 
With the focus shifting from more precise constraints of the $\Lambda$CDM parameters to testing new physics beyond the standard cosmological model or resolving cosmological tensions, more and more cosmological parameters are introduced, which can have particularly complex parameter structures complicating the parameter inference. These developments motivate increased care in the statistical analysis to avoid loss of constraining power and, more importantly, to avoid that cosmological constraints are influenced by unwanted or unknown effects of the statistical analysis. 

The statistical toolkit available to cosmologists encompasses a variety of methods,
for example, the Bayesian and the frequentist frameworks.\footnote{There are other frameworks, including hybrid methods that combine elements of Bayesian and frequentist statistics}
%, e.g.\ \cite{Amendola:2024prl}.}. 
Each of these tools has advantages and disadvantages and using different methods can help detect and mitigate statistical effects that can lead to misleading results in the analysis. We see this interplay being used in fields like particle physics where these two frameworks are often used to better understand the data and optimize information extraction from experiments (e.g.\ \cite{Trotta:2008bp, Kowalska:2012gs, Lyons:2013ch, BaBar:2017tiz, Agostini:2017iyd}). 

In the predominantly used Bayesian framework, the main object is the posterior probability constructed from the likelihood and the assumed prior on the model parameters~\cite{Verde:2009tu, Trotta:2017wnx}. In cases where the data is not constraining, the likelihood surface will be flat and the posterior can have a strong dependence on the chosen prior. Further, the Bayesian framework deals with nuisance parameters by marginalizing over them in the multidimensional posterior. This leads to a built-in dependence of the inferred parameter intervals on the multidimensional prior volume, also called (prior) volume or projection effect, which up-weights regions with larger volume regardless of the quality of the fit. While this dependence can be viewed as a feature since it reflects the preference for the larger parameter volume, a strong impact of prior volume effects is often unwanted. 
In cases, where the priors are not well motivated, this sensitivity of the parameter intervals on the chosen prior can be particularly undesirable. Moreover, parameters targeted by upcoming surveys, like the sum of neutrino masses or the tensor-to-scalar ratio, are close to physical boundaries, which can lead to a bias towards values further away from the boundary, resulting in an over- or underestimation of uncertainties\footnote{The truncation of the part of the posterior beyond the boundary can lead to an underestimation, while artificially enforcing exploration of parameters further away from the boundary can lead to an overestimation of the uncertainty.}. Proximity to a boundary can also lead to a dependence on the choice of lower limit in the prior as well as to the shape of the prior, e.g.\ linear vs.\ logarithmic prior. A strong sensitivity of the parameter constraints on the prior was reported in the cosmology literature in different contexts, e.g. neutrino mass bounds~\cite{Gonzalez-Morales:2011tyq, Planck:2013nga, Simpson:2017qvj, Gariazzo:2018pei,Gariazzo:2023joe,DESI:2024mwx,Naredo-Tuero:2024sgf,Craig:2024tky,Green:2024xbb}; early dark energy~\cite{Smith:2020rxx, Gsponer:2023wpm}, full-shape galaxy clustering with the effective field theory of large scale structure ~\cite{Carrilho:2022mon,Moretti:2023drg, Donald-McCann:2023kpx, Holm:2022kkd}, and stage-IV forecasts~\cite{Hadzhiyska:2023wae}.

In frequentist statistics, the main object is the likelihood. There is no built-in dependence on priors\footnote{Although prior information can also be used in frequentist statistics by building a joint likelihood of current and prior experiments.} and the likelihood is parametrization invariant, which makes it insensitive to the problems above. This lack of dependence on priors makes frequentist constraints especially useful in situations when the inferred parameter interval has a strong dependence on the prior, when prior volume effects are dominating the constraints, or when the cosmological limits are close to a physical boundary. In those cases, the frequentist intervals can provide important complementary information about the analysis, in particular when used along the Bayesian analysis to explore the effects above.

Frequentist methods, in particular, profile likelihoods have been applied in several cosmological settings, e.g.\ to verify the Bayesian constraints on the $\Lambda$CDM parameters from \Planck~\cite{Planck:2013nga} cosmic microwave background (CMB) data, to determine the baryon acoustic oscillation (BAO) scale from galaxy surveys~\cite{BOSS:2012tck,eBOSS:2017cqx,DES:2017rfo,DES:2018fiv,Ruggeri:2019kjl,Cuceu:2020dnl}; to constrain evolving dark energy~\cite{Yeche:2005wn}, extra number of relativistic species, $N_\mathrm{eff}$ \cite{Hamann:2007pi,Hamann:2011hu,Henrot_Versill__2019}, cosmic strings \cite{Henrot_Versill__2015}, the sum of neutrino masses \cite{Reid:2009nq,Planck:2013nga,Couchot_2017,Gonzalez-Morales:2011tyq, eBOSS:2020yzd, Giare:2023qqn,Naredo-Tuero:2024sgf, Noriega:2024lzo} or in the context of $\Lambda$CDM+$A_L$ \cite{Couchot_2017_2}.
Recently, renewed interest in frequentist methods has appeared in the context of the Hubble tension for the early dark energy model \cite{Herold:2021ksg, Herold:2022iib,Reeves:2022aoi,Efstathiou:2023fbn} and new early dark energy models~\cite{Cruz:2023cxy}; for other beyond-$\Lambda$CDM models like decaying dark matter \cite{Holm:2022kkd}, phenomenological transition from dark matter to dark radiation~\cite{Holm:2023uwa,Bringmann:2018jpr}, coupled quintessence and modified gravity \cite{Gomez-Valent:2022hkb}; and neutrino-dark matter interactions~\cite{Giare:2023qqn}; or to study the effect of priors on nuisance parameters of the effective field theory of large scale structure \cite{Holm:2023laa}; in inflation~\citep{Campeti:2022acx,LiteBIRD:2023zmo,Galloni:2024lre} the tensor-to-scalar ratio~\citep{Campeti:2022vom,SPIDER:2021ncy,Galloni:2024lre,Capistrano:2024kuc}; and in measurements of the expansion history~\citep{Colgain:2023bge, Colgain:2024clf}; see also~\cite{Wagner:2018jxp, Kerscher:2024doc} for alternative (Bayesian) frequentist approaches. 

However, frequentist methods also have their own shortcomings: they can prefer cosmologies with very small parameter volumes since they are insensitive to the parameter volume (``fine-tuning''). Further, computing frequentist confidence intervals with the full Neyman construction is computationally expensive since it requires the evaluation of the likelihood for many mock data sets. While in the asymptotic limit, the simpler graphical profile likelihood method can be used, it is still computationally expensive compared to MCMC in cases with many parameters due to many minimizations in large-dimensional spaces. It is not always clear that the profile likelihood method can be used, i.e.\ that the asymptotic limit is reached such that proper frequentist coverage\footnote{A confidence interval is said to cover if -- upon repetition of the experiment -- the true value of the parameter is contained within a fraction $(1-\alpha)$ of the constructed intervals as required by the C.L., $(1-\alpha)$.} is guaranteed. If intervals are near a physical boundary, a more elaborate construction needs to be used (``Feldman-Cousins construction''; see Sec.~\ref{sec:Gaussian_boundaries}).
Hence, as in the Bayesian case, care needs to be taken to quote meaningful confidence intervals.

While Bayesian methods have been largely studied in the literature on cosmology, frequentist methods have not been widely used in cosmology and therefore, the literature on the topic is scarce. In this context, we aim to give a pedagogical introduction of profile likelihoods and general frequentist confidence intervals in cosmology. Most importantly, the graphical profile likelihood construction relies on assumptions that are rarely checked and, therefore, in this paper, we aim to present, for the first time to our knowledge, a detailed analysis of the validity of the assumptions when constructing these confidence intervals.

The motivation for this paper is threefold: We discuss \textit{why} and in which situations it is advantageous to compute profile likelihoods in cosmology. We describe \textit{when} the graphical profile likelihood construction gives meaningful confidence intervals with correct coverage. Finally, we review \textit{how} to obtain confidence intervals with the profile likelihood method.
Our main goal is to assess the validity of the assumptions necessary for the graphical profile likelihood method to give correct coverage, including different complicated situations in cosmology like the presence of physical boundaries and/or situations where Wilks’ theorem does not hold. We evaluate this for different cosmological models $\Lambda$CDM, $\Lambda$CDM with the total sum of the neutrino masses, $M_\nu$, as a free parameter, and $\omega_0$CDM, where the equation of state of dark energy is a free parameter, using \textit{Planck} 2018 \texttt{Plik\_lite} data. 

This paper is organized as follows: Sect.~\ref{sec:cookbook} provides a profile likelihood ``cookbook,'' which is meant for the reader mainly interested in the practical use of profile likelihoods in cosmology. Sec.~\ref{sec:freq_constraints} gives a more detailed pedagogical review of frequentist confidence intervals. 
In Sec.~\ref{sec:pinc}, we introduce \pinc, a simple code for computing profile likelihoods in cosmology. Sec.~\ref{sec:data_and_mocks} details the data sets and the generation of mock \textit{Planck}-lite data. 
In Sec.~\ref{sec:results} we probe the validity of the asymptotic assumptions such as Wilks' theorem in $\Lambda$CDM, $\Lambda$CDM$+M_{\nu}$, and $\omega_0$CDM. Finally, we report frequentist and Bayesian intervals for the three models under CMB and BAO data in Sec.~\ref{sec:profiles} and conclude in Sec.~\ref{sec:conclusions}.

%%%%%%%%%%%%%%%%%%%%%%%%%%%%%%%%%%%%%%%%%%%%%%%%%%%%%%%%%%%%%%%%%%%%%%

\section{Profile likelihood cookbook}
\label{sec:cookbook}

This section is to guide the reader who is only interested in the practical application of profile likelihoods. To this end, we discuss frequently asked questions about why, when, and how to compute frequentist confidence intervals.

\subsection{Why compute frequentist confidence intervals?}

In certain circumstances, it can be interesting to compute a frequentist interval in addition to a Bayesian interval. Since Bayesian posteriors are simple and (relatively) cheap to obtain via MCMC, we assume that Bayesian constraints are already available for the model of interest. If some of the parameters of the model are not well constrained or at a physical boundary, it is important to (1) assess the sensitivity of the results on the choice of prior and/or (2) compute point estimates like the maximum likelihood estimate (MLE) or maximum \textit{a posteriori} (MAP). If (1) one finds a dependence of the results on the choice of prior and/or (2) finds that the MLE/MAP strongly deviates from the mean, e.g.\ is outside of the $1\sigma$ credible interval, this points to a (possibly unwanted) impact of the prior or prior volume effects. If this is the case, it is interesting to compute a frequentist confidence interval, which is inherently prior independent and can be used to probe the impact of these effects.

\subsection{When does the graphical profile likelihood construction give constraints with correct coverage?}

The most general procedure to construct frequentist confidence intervals is the Neyman construction, which guarantees correct coverage. However, since it is computationally expensive, this construction is usually avoided and the approximate graphical profile likelihood construction is used instead. 
The graphical profile likelihood method gives correct coverage for a Gaussian parameter distribution or \textit{in the limit of a large data set} (Wilks' theorem, Sec.~\ref{sec:wilks}).
Checking whether Wilks' theorem holds in practice is often infeasible. We probe the validity of Wilks' theorem for \Planck-lite data in Sec.~\ref{sec:results} to find indications for its validity. 

Note that it is \textit{not} sufficient if the profile likelihood represents a parabola. A parabolic profile likelihood shows only that the likelihood for the observed data is Gaussian. For correct coverage, however, the likelihood needs to be Gaussian for all choices of model parameters and (hypothetically observed) data. The $w_0$CDM model under \Pliklite\ data is an example, where the profile likelihood is well described by a parabola near the MLE (Fig.~\ref{fig:PL_w0}) but our mock tests indicate that Wilks' theorem does not hold (bottom panel of Fig.~\ref{fig:LR_hist_w0}).

\subsection{How to compute the profile likelihood and frequentist confidence intervals?}

Once one has decided to compute a frequentist confidence interval, one needs to decide whether to use the time-consuming full Neyman construction or the simpler graphical profile likelihood method. If generating mock data and evaluating the likelihood is fast, one can consider conducting a full Neyman construction (as in e.g.\ \cite{SPIDER:2021ncy, LiteBIRD:2023zmo}). However, if the evaluation of the likelihood is expensive, the only feasible option might be the graphical profile likelihood method. In this case, it is common to assume that the Gaussian approximation or Wilks' theorem holds (Sec.~\ref{sec:wilks}), and the graphical profile likelihood method is used while acknowledging that correct coverage might not be fulfilled. The necessary steps for constructing frequentist confidence intervals under the Gaussian approximation or given that Wilks' theorem holds, are: 
\begin{itemize}
    \item[1.] Compute a profile likelihood using an efficient minimizer. One can use, for example, one of the public codes referenced in Sec.~\ref{sec:pinc}, including our code \lowercase{\pinc}.
    \item[2.] Construct a confidence interval. For that, it is relevant whether the parameter is near a physical boundary. If there is no boundary, one can use the simple graphical profile likelihood method based on iso-likelihood contours (Sec.~\ref{sec:Gauss_PL}). If the parameter is near a physical boundary, one needs to use the boundary-corrected graphical construction (Sec.~\ref{sec:Gaussian_boundaries}).
\end{itemize}
We review frequentist parameter inference in some detail in the next section. The reader, interested in the results, can skip to Sec.~\ref{sec:results}.

%%%%%%%%%%%%%%%%%%%%%%%%%%%%%%%%%%%%%%%%%%%%%%%%%%%%%%%%%%%%%%%%%%%%%%

\section{Construction of frequentist parameter constraints}
\label{sec:freq_constraints}

\subsection{Setting the stage: Bayesian credible intervals}

In Bayesian statistics -- which is commonly used in cosmology -- one associates a probability to the model parameters. The key quantity is the posterior $P(\allpar|\data)$, which gives the probability of the model parameters $\allpar$ given the data $\data$.\footnote{We omit the dependence of the posterior $P(\allpar|\data, M)$ and all other quantities on the model, $M$, for conciseness.} The posterior can be related to the likelihood $\like_{\data}(\allpar) = P(\data|\allpar)$ via Bayes theorem,\footnote{We write $\data$ as subscript to emphasize that the likelihood $\like_{\data}(\allpar)$ is a function of the parameters, $\theta$, for a fixed data, $\data$.}
\begin{equation}
    P(\allpar|\data)
    \sim \like_{\data}(\allpar)\cdot \pi(\allpar),
\end{equation}
where $\pi(\allpar)$ is the prior, which represents the prior beliefs about the model parameters and has to be picked by the data analyst. If nuisance parameters are present, we split the parameter space into parameters of interest (e.g. cosmological parameters) $\poi$ and nuisance parameters $\np$, $\allpar = (\poi, \np)$.
If the model does not only contain (cosmological) parameters of interest, $\poi$, but also nuisance parameters, $\vec\nu$, one obtains the posterior of the parameters of interest via marginalization, i.e.\ integration over the nuisance parameters:
\begin{equation}
    P(\poi|\vec x) = \int P(\poi, \np| \vec x)\ \d\np.
\end{equation}
In practice, the above integral does not need to be solved explicitly but one can easily obtain the marginalized posterior from a sample of the full posterior by simply disregarding the parameter to be marginalized over. For a probability $(1-\alpha)$, the credible interval $[\onepoi_1,\onepoi_2]$ for a parameter $\onepoi$ is obtained via integration of the posterior:
\begin{equation}
    \label{eq:Bayesian_interval}
    P(\onepoi \in [\onepoi_1,\onepoi_2]|\data) = \int_{\onepoi_1}^{\onepoi_2} P(\onepoi|\data) \d \onepoi = 1-\alpha.
\end{equation}
The interval $[\onepoi_1, \onepoi_2]$ can be chosen e.g.\ as a central interval, i.e.\ $P(\onepoi \leq \onepoi_1|\data) = P(\onepoi \geq \onepoi_2|\data) = \frac{\alpha}{2}$, or as an upper/lower limit, i.e.\ $P(\onepoi \geq \onepoi_{2/1}|\data)=\alpha,\ P(\onepoi \leq \onepoi_{1/2}|\data) = 0$. Bayesian intervals assign a probability to the value of the (model) parameter $\onepoi$. The interpretation of the interval in Eq.~\eqref{eq:Bayesian_interval} could be phrased as: ``The degree of belief that the true value of the parameter $\onepoi$ lies in $[\onepoi_1,\onepoi_2]$ is $(1-\alpha)$ given the observed data and my prior beliefs''. Thus Bayesian intervals are also called ``credible intervals.''

\subsection{Neyman construction}
\label{sec:Neyman_constr}

\begin{figure}
    \centering
    \subfigure{\includegraphics[scale=0.55]{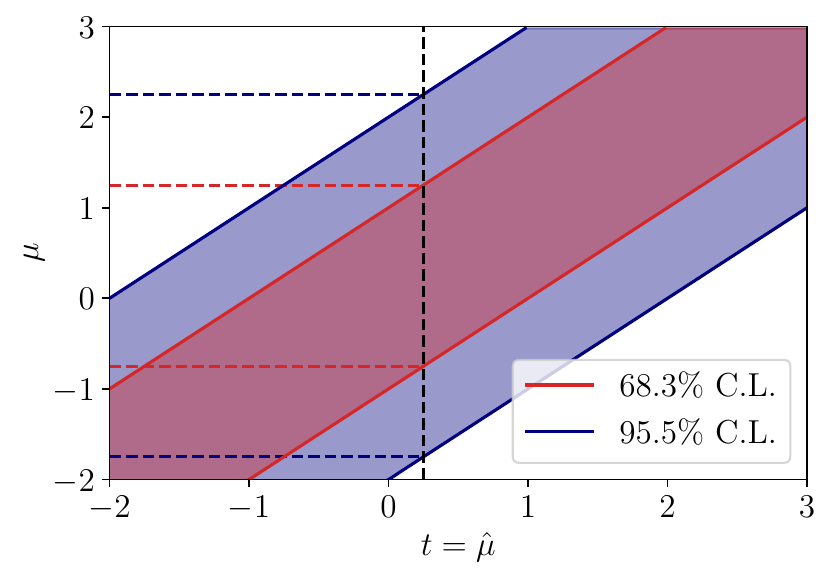}}
    \includegraphics[scale=0.55]{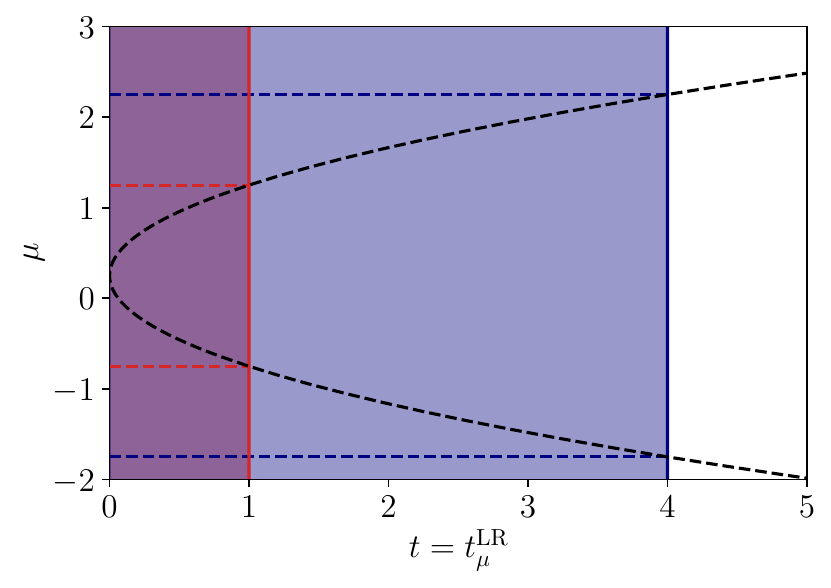}
    \includegraphics[scale=0.55]{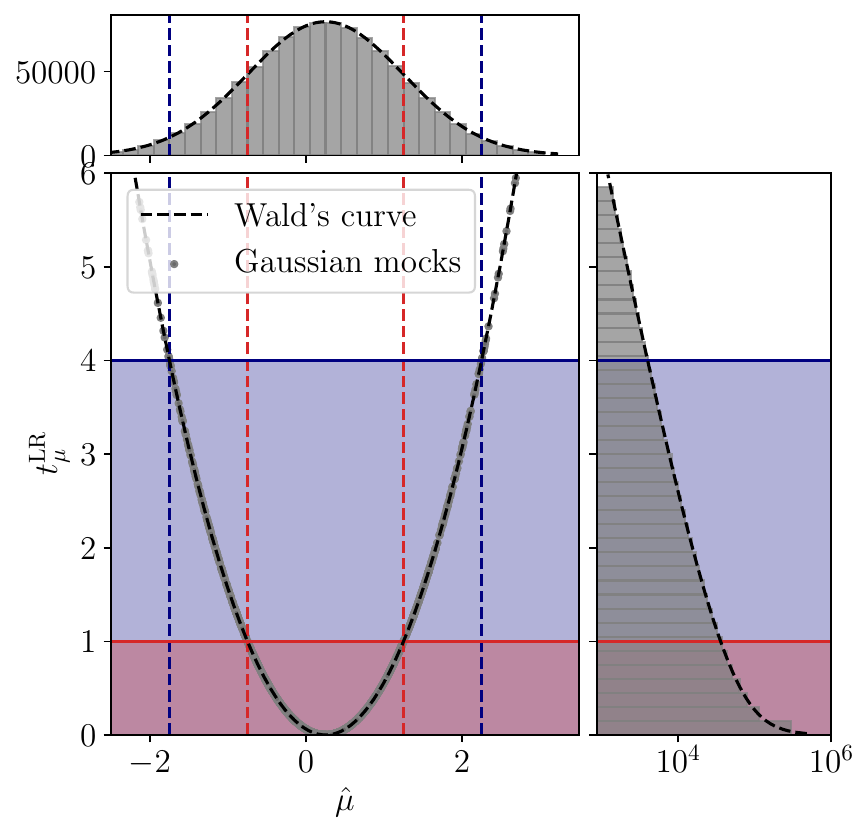}
    \caption{Neyman confidence belts for a Gaussian parameter $\onepoi$ at $68\%$ ($95\%$) C.L.\ are shown as the red (blue) shaded bands with the test statistic $t=\hat\onepoi(\data)$ (\textit{top}) and $t=t_\onepoi^\LR(\data)$ (Eq.~\ref{eq:tLR_Gauss}, \textit{center}). Upon measuring a value of $t_\obs = 0.25$, the confidence interval of the model parameter $\onepoi$ (red and blue dashed horizontal lines) is given by the intersection of $t = t_\obs$ (black dashed line) with the Neyman belt. 
    \textit{Bottom:} mocks drawn from a Gaussian distribution (grey markers), showing $t^\mathrm{LR}_\onepoi(\data)$ as a function of the inferred $\hat\onepoi$, which follow Wald's curve (black dashed line). The red (blue) shaded regions contain $68\%$ (95\%) of the mocks as shown in the histograms along the $x$ and $y$ axes. This illustrates why the graphical profile likelihood construction gives correct coverage for a Gaussian distribution (see text).}
    \label{fig:Neyman_Gauss}
\end{figure}

In frequentist statistics on the other hand, ``confidence intervals'' are constructed (see \cite{Cousins:2024gkj} for a review). Instead of assigning a probability to the model parameters, intervals are designed such that they have well-defined frequentist properties under repeated experiments. The central property is \textit{coverage}: an interval estimation procedure is said to cover if the true value of the parameter $\onepoi_\true$ lies within a fraction $(1-\alpha)$ of the constructed intervals, where $(1-\alpha)$ is the confidence level (C.L.)\footnote{In fact, Bayesian intervals derived from $P(\allpar|\data)$ are also random objects since they depend on random data $\vec x$. Hence, one can also study coverage for Bayesian intervals but it is not their defining property and not of high priority in the Bayesian setting.}.

A confidence interval with exact coverage can be created using the \textit{Neyman construction} \cite{Neyman:1937uhy}, which is based on a simple idea: construct a hypothesis test of size $\alpha$ for all values of $\onepoi$ and define the interval for some observed data $\data$ as the set of hypotheses which are not rejected by that data. This procedure is often referred to as an \emph{inversion of hypothesis tests}. The correct coverage is evident: if the data originated from a true parameter value $\onepoi$, it would not be rejected by that parameter's hypothesis test --- and thus be included in the constructed interval --- $(1-\alpha)$ of the time. We discuss the construction in more detail below.

To define the hypothesis tests for a hypothesis given by a parameter choice $\allpar$, we make use of a scalar test statistic $t(\data)$, which is a function of the data $\vec x$. A common choice is to simply use an estimator of the parameter of interest: $t(\data)= \hat\onepoi(\data)$, e.g.\ the MLE or best fit. But it is worth noting that the specific choice of test statistic may additionally depend on the parameters $\allpar$: $t(\vec x) = t_{\allpar}(x)$. To create a well-defined hypothesis test, we must know the density $p(t|\allpar)$ of the test statistic $t$ given the parameters $\allpar$. While in some cases $p(t|\allpar)$ is known in closed form, in general it is not. However, this density can always be estimated by e.g. sampling from $x_i \sim p(\data|\allpar)$ and histogramming the values $t(x_i)$ as follows: for one fixed choice of $\allpar$, one generates many mock realizations $\data_i$; for each of these mock realizations $\data_i$, one obtains an estimate of the test statistic, $\hat t_{\allpar}(\data_i)$. This allows to approximate the distribution $p(t|\allpar)$ of the test statistic $t$ given $\allpar$. 

Given a density $p(t|\allpar)$, one can then use an \emph{ordering rule} to define an \emph{acceptance region} $\mathcal{T}_{\allpar}$.
Common ordering rules are for example a central interval, i.e. $p(t<t_1|\allpar) = p(t>t_2|\allpar) = \frac{\alpha}{2}$, or an upper limit, i.e.\ $p(t>t_2|\allpar) = \alpha$. Given $\mathcal{T}_{\allpar}$, one can set up a hypothesis test for the parameter choice $\allpar$: the hypothesis is \emph{rejected} if the test statistic of the observed data $\data$ falls outside of this region. The region is chosen such that the probability of rejecting $\data$ originating from $p(\data|\allpar)$ has a known rate $\alpha$: 
\begin{equation}
     \label{eq:Neyman_belt}
     p(t \in \mathcal{T}_{\allpar}|\allpar) 
     = \int_{\mathcal{T}_{\allpar}} p(t|\allpar)\ \d t
     = 1-\alpha.
\end{equation}
The boundaries of these regions often vary smoothly as a function of the parameter and one can think of the union of all acceptance regions as a ``Neyman belt.'' The interval is now constructed by plotting the observed test statistic $t_{\allpar}(\vec x_\mathrm{obs})$ as a function of the parameter values. The interval is then defined as the region where the \emph{observed} test statistic $t_{\allpar}(\data_\obs)$ lies within the belt region.

\subsection{Gaussian model and the graphical method}
\label{sec:Gauss_PL}

We illustrate the procedure with a simple Gaussian example $p(t|\onepoi) = \mathcal{N}(t|\onepoi,\sigma)$ with a single parameter of interest $\onepoi$, a fixed standard deviation $\sigma$ and no nuisance parameters.\footnote{This is of particular interest as it corresponds to the asymptotic limit of a model $p(\data|\onepoi)$ and a choice of the MLE estimator as the test statistic $t(\data)=\hat{\onepoi}(\data)$ (Sec.~\ref{sec:wilks}).} The test statistic is, therefore, $t(\data) = \hat{\onepoi}(\data)$.
Recall that asymptotically the MLE estimates are unbiased and distributed normally around the true value $\onepoi$ with a variance defined by the inverse Fisher information. In this case, a natural approach to define the acceptance regions $\mathcal{T}_\onepoi$ is to define a central interval $[t_1,t_2]$ such that the left and right tail each hold $\alpha/2$ of probability mass. As the test statistic distribution is centered on the true value $\onepoi$, the boundaries of this central interval move to the right as $\onepoi$ is increased.
This produces a ``Neyman belt'' that lies diagonally across the $(t,\onepoi)$ plane as shown in the top panel of Fig.~\ref{fig:Neyman_Gauss} as the red and blue shaded regions. As the test statistic $t(\data) = \hat{\onepoi}(\data)$ does not depend on $\onepoi$, it is constant as a function of the parameter $\onepoi$ and thus corresponds to a vertical line in the $(t,\onepoi)$ plane cutting through the ``Neyman belt'' (black dashed line in the top panel of Fig.~\ref{fig:Neyman_Gauss}). The interval $[\onepoi_1, \onepoi_2]$ starts where the vertical line enters the belt from below at $\onepoi = \onepoi_1$ and ends at $\onepoi=\onepoi_2$ where it exits it again (red and blue dashed horizontal lines).

There is an intimate connection~\cite{Cranmer:2014lly} of this Neyman construction in the Gaussian case to another popular interval construction technique: the graphical method. Consider an alternative test statistic
\begin{equation}
    \label{eq:tLR_Gauss}
    t^\LR_\onepoi(\data) = -2\log  \frac{p(t|\onepoi)}{p(t|\hat{\onepoi})} = \frac{(t-\onepoi)^2}{\sigma^2}.
\end{equation}
In a simple Gaussian setting and for an observed value $t$, the value $\hat{\onepoi}$ that maximizes the likelihood is simply $\hat{\onepoi} = t$ and the log-likelihood ratio (LR) above takes on a simple parabolic form. Note that now the choice of test statistic \emph{varies} as a function of the parameter $\onepoi$. 

As per the recipe, we need to think of the distribution $p(t^\LR_\onepoi|\onepoi)$. Since we assume that $p(t|\onepoi)$ is distributed according to a Gaussian distribution centered at $\onepoi$, we can deduce the distribution of its transformation $t\to t^\LR_\onepoi = (t-\onepoi)^2/\sigma^2$ easily: Gaussian random variables distributed around some mean mapped through a parabola anchored at the same mean are distributed according to the $\chi^2$ distribution irrespective of what the value of the mean is. Therefore,
$$p(t^\LR_\onepoi|\onepoi) = \chi^2\;\,, \, \forall \onepoi.$$
Thus unlike the previous case, the distribution of the test statistic is now \emph{constant} for all values $\onepoi$. This can be understood as the result of two changes that cancel each other out: as we change $\onepoi$ the distribution $p(t|\onepoi)$ changes. But at the same time the test statistic we consider $(t-\onepoi)^2/\sigma^2$ changes as well. Together these two changes yield a static distribution $p(t^\LR|\onepoi)$. 

Continuing with the construction, we can define acceptance regions. Here, high values of $t_\onepoi^\LR$ correspond to large deviations of $t$ from the central value $\onepoi$. The analogue of the central region in $t$ would thus be to define the acceptance regions such that $t^\LR < t_0$, where $t_0$ is a cutoff value such that the test has the desired size $\alpha$. For standard test sizes and the $\chi^2$ distribution, these are the familiar cutoff values $t_0 = 1,4,9,\dots$. The acceptance regions are thus independent of $\onepoi$, $\mathcal{T}_\onepoi = \mathcal{T}$, and the ``Neyman belt'' is just a fixed vertical band at the corresponding cutoffs as can be seen in the center panel of Fig.~\ref{fig:Neyman_Gauss}. 

The last step of the construction is to draw the observed data in the $(t^\LR,\onepoi)$ plane. For some observed data value of the original test statistic $t_\obs$, our new test statistic now varies as a function of $\onepoi$ and thus is no longer a straight vertical line. Instead, it resembles a parabola, where the value of $t^\LR_\onepoi$ vanishes at $\onepoi = t$ (black dashed line in the center panel of Fig.~\ref{fig:Neyman_Gauss}). The interval is the set of parameter values where this parabola lies below the $\chi^2$ cutoff values (red and blue dashed horizontal lines). 

This is nothing else than the ``graphical method'' of confidence intervals in disguise. Recall that in the graphical method one plots the log-likelihood curve normalized to the MLE value $t_\onepoi^{LR}(\data) =-2\log \frac{p(t|\onepoi)}{p(t|\hat{\onepoi})}$ and defines the interval as the $\onepoi$ range where that curve stays below a $\chi^2$ cutoff value of $1,4,9\dots$ for 68\%, 95\%, 99\% C.L., respectively.

\subsection{Nuisance parameters and profile likelihood}

Crucially, the simple results of the Gaussian model generalize for models with nuisance parameters, $p(\data|\poi,\np)$. As discussed above the MLE estimators become asymptotically Gaussian
%within the asymptotic limit 
and one can think of one choice of test statistic, $t=\hat{\onepoi}(\data)$, with a variance inversely proportional to the Fisher information. 

Separately, one can extend the LR definition, Eq.~\eqref{eq:tLR_Gauss}, to the standard \textit{profile likelihood ratio} test statistic:
\begin{equation}
    \label{eq:t_PLR}
     t_{\poi}^{\LR}(\data) = - 2 \log\frac{\like_{\data}(\poi, \hhat{\np})}{\like_{\data}(\hat\poi,\hat{\np})},
\end{equation}
where $\hat\poi$ and $\hat{\np}$ denote the MLE estimators of the parameters of interest and nuisance parameters, while $\hhat{\vec\nu}$ denote the ``conditional'' MLE estimators for $\np$ obtained from a constrained optimization where $\poi$ are kept fixed. It is notable that even in the presence of nuisance parameters the test statistic, Eq.~\eqref{eq:t_PLR}, only depends on the parameters of interest, $\poi$. Hence, while Bayesian statistics handle nuisance parameters through marginalization, frequentist statistics do so by optimization instead.

\subsection{Asymptotic theory and Wilks' theorem}
\label{sec:wilks}

The correspondence between the graphical method and the Neyman construction with a test statistic based on the LR extends well beyond the simple model of a Gaussian but holds for any statistical model $p(\data|\allpar)$ that is within the asymptotic regime of a large data set. This is due to the fact that many relations simplify and become Gaussian within the asymptotic limit. 

A major result is \emph{Wilks' theorem}~\cite{Wilks:1938dza} that generalizes the result from Sec.~\ref{sec:Gauss_PL}: in the asymptotic limit, i.e.\ in the limit of a large data set, the distribution of the (profile) LR test statistic, Eq.~\eqref{eq:t_PLR}, takes on a fixed form and is $\chi^2$-distributed. Moreover, for alternative hypotheses, i.e.\ values of $\onepoi$, which are different from the true value $\onepoi_\true$, the test statistic in Eq.~\eqref{eq:t_PLR} follows a \textit{non-central} $\chi^2$ distribution.

The results of asymptotic normality and Wilks' theorem are tied together by another asymptotic result by Wald~\cite{Wald:1943} (which we refer to as \textit{Wald's relation/curve}) that connects the maximum likelihood estimate of the parameter of interest, $\onepoi$, with the profile likelihood in the asymptotic regime:
\begin{equation}
    \label{eq:Walds_relation}
    t_{\onepoi}^\LR(\data) = \frac{(\hat{\onepoi}(\data)-\onepoi)^2}{\sigma_{\onepoi}^2}.
\end{equation}
This relation is illustrated in the bottom panel of Fig.~\ref{fig:Neyman_Gauss} for a Gaussian parameter $\onepoi$ with true value $\onepoi_\true = 0.25$ and standard deviation $\sigma_{\onepoi} =1$. Realizations drawn from $\hat\mu_i \sim \mathcal{N}(\hat\onepoi|\onepoi,\sigma_\onepoi)$ (grey markers) lie on the curve described by Eq.~\eqref{eq:Walds_relation} (black dashed line). The LR test statistic, $t_{\onepoi}^\LR(\data)$, Eq.~\eqref{eq:tLR_Gauss}, is distributed as $\chi^2$ (histogram along the $y$ axis). From this, it is easy to see why the graphical method gives correct coverage in the case of a Gaussian: $68\%$ ($95\%$) of the such generated mocks lie below the familiar $\chi^2$ cutoff-values 1 (4). 
The standard deviation $\sigma_\mu$ in Wald's relation is often estimated from the so-called \textit{Asimov data set} \cite{Cowan:2010js}, which refers to a mock realization of the data with all model parameters fixed to the ground truth (see App.~\ref{sec:Asimov}).

It is remarkable that in this asymptotic regime all three ingredients 1) the asymptotic normality of the MLE estimators, 2) the distribution of the profile LR test statistic (Wilks' theorem), and 3) the relationship between the two (Wald's relation) \emph{do not} depend on the nuisance parameters or the details of the model $p(\data|\allpar)$.

This has a profound effect on confidence intervals: according to the Neyman construction, one would normally have to estimate the distribution $p(t|\poi,\np)$ of the test statistic for all possible combinations of parameters of interest and nuisance parameters, $(\poi, \np)$, which can become intractable for many nuisance parameters. If $p(t_{\poi}^\mathrm{LR}|\poi,\np) = p(t_{\poi}^\mathrm{LR}|\poi)$, it is sufficient to carry out the construction purely in the space of the parameters of interest, which in turn is very simple: within the asymptotic regime, the Neyman construction simplifies to the graphical construction, i.e.\ just considering the iso-contours of the profile likelihood and declaring them as confidence intervals.

The simplifications the asymptotic theory affords are so significant, that 
%the techniques for interval construction often used even 
the prerequisites, i.e.\ that a model $p(\data|\allpar)$ is well within the asymptotic regime, are often not checked in detail. However, we point out that only then does, e.g. the graphical method, yield correct coverage. In cases that are less regular, the procedure must be adapted accordingly.

\subsection{Physical boundaries and Feldman-Cousins}
\label{sec:Gaussian_boundaries}

\begin{figure}
    \centering
    \subfigure{\includegraphics[scale=0.55]{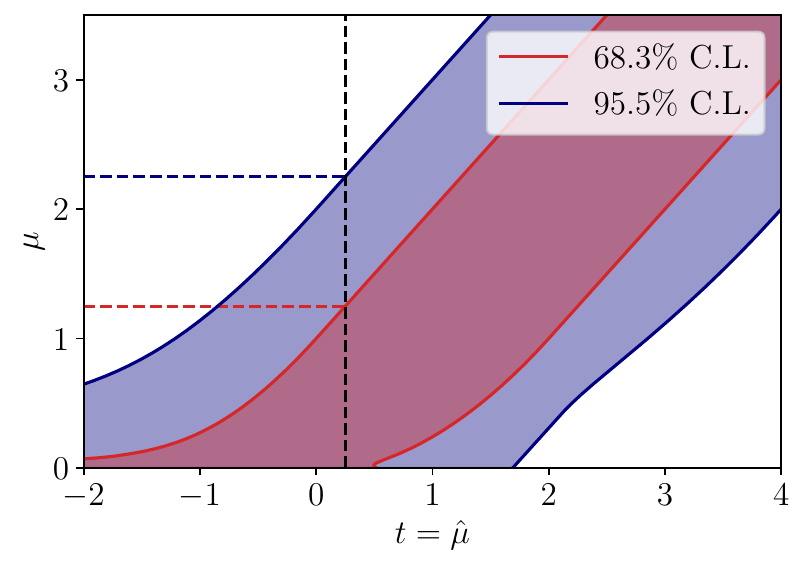}}
    \includegraphics[scale=0.55]{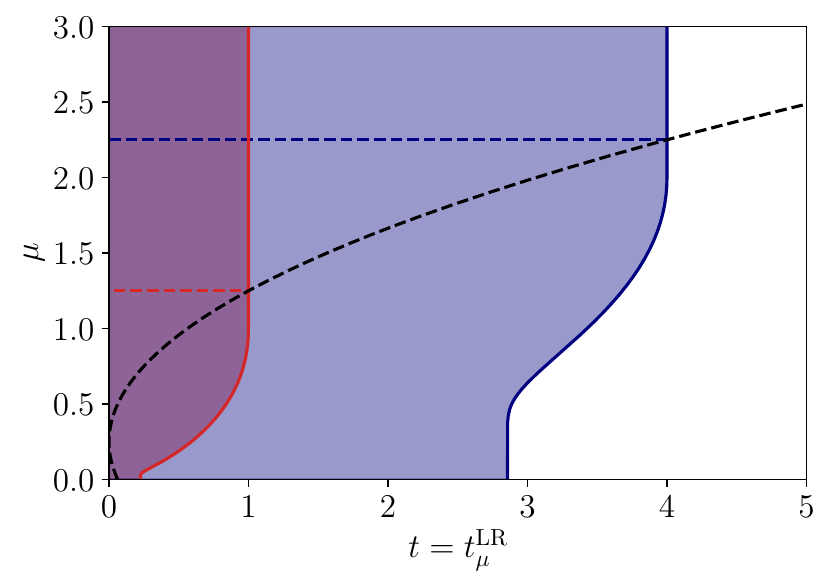}
    \includegraphics[scale=0.55]{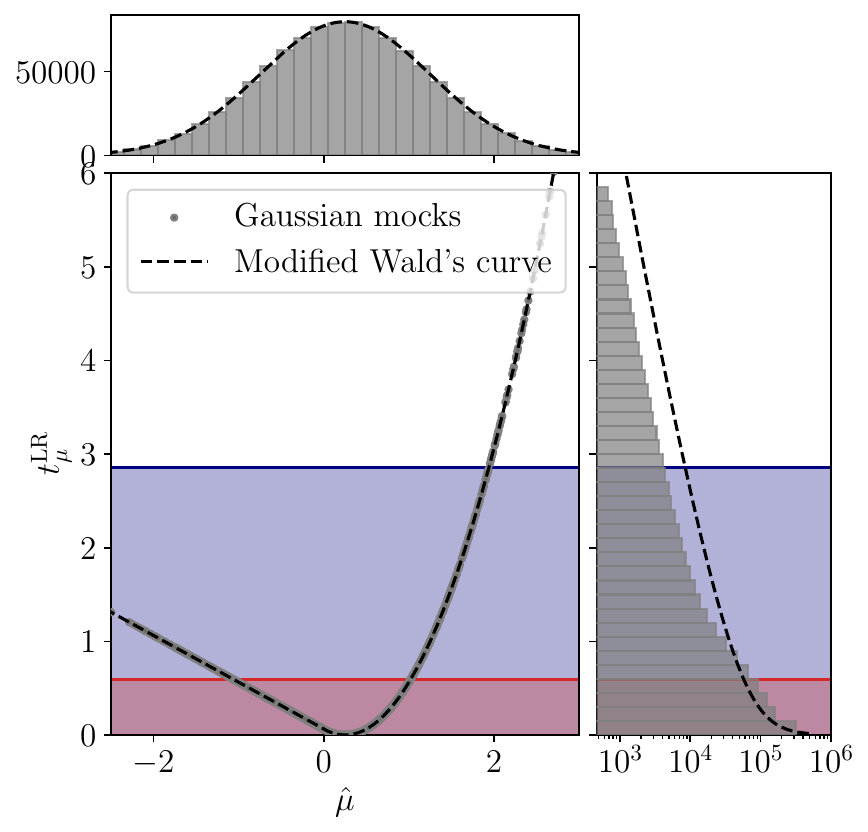}
    \caption{Neyman confidence belt for a Gaussian parameter $\onepoi$ near a physical boundary in $\onepoi=0$. The boundary leads to a modification of the acceptance regions near $\onepoi=0$ for the test statistic $t=\hat\onepoi(\data)$ with ordering rule $t^\LR_\onepoi < t_0$ (Feldman-Cousins, \textit{top}) and $t=t_\onepoi^\LR(\data)$ (Eq.~\ref{eq:tLR_boundary}, \textit{center}). The interval defined by the intersection of $t = t_\obs$ (black dashed line, here $t_\obs = 0.25$) and the Neyman belt smoothly transitions between upper/lower limits and central intervals.  
    \textit{Bottom:} the boundary leads to a mixed distribution, Eq.~\eqref{eq:mod_Wald}, which is linear for $\hat\onepoi<0$ and follows a parabola for $\hat\onepoi>0$ (black dashed line). The distribution of $t_\onepoi^\LR(\data)$ (histogram along $y$ axis) deviates from $\chi^2$ (black dashed line), i.e.\ Wilks' theorem is violated.}
    \label{fig:Neyman_boundary}
\end{figure}

A first deviation from the graphical method is necessary when considering physical boundaries such as  $\poi \geq 0$ -- but still assuming asymptotic behavior, as here Wilks' theorem does \textit{not} hold. In the presence of a boundary, Wald's relation, which says that the MLEs $\hat{\poi}$ and the test statistic are parabolically related (Eq.~\ref{eq:Walds_relation}), does not hold anymore and thus the $\chi^2$ distribution for the test statistic from Wilks' theorem also breaks down. We review this situation in the one-dimensional case with one parameter of interest, $\onepoi$, below.

In this case, the construction described in Sec.~\ref{sec:Gauss_PL} can be adapted as follows. Global optimization in the denominator of the profile LR optimizes for the best \emph{physical} value $\hat{\onepoi}_\mathrm{phys}$, where for $\hat{\onepoi} < 0$, the denominator is replaced by the likelihood value \emph{at} the boundary. Thus, Eq.~\eqref{eq:t_PLR} becomes
\begin{equation}
    \label{eq:tLR_boundary}
    t_{\onepoi}^\LR(\data) = \begin{cases} - 2 \log\frac{\like_{\data}(\onepoi, \hhat{\vec\nu})}{\like_{\data}(0,\hhat{\np}_0)}\;\mathrm{if}\; \hat{\onepoi} < 0,\vspace*{0.2cm}\\
     - 2 \log\frac{\like_{\data}(\onepoi, \hhat{\vec\nu})}{\like_{\data}(\hat{\onepoi},\hat{\np})}\;\mathrm{otherwise},
    \end{cases}
\end{equation}
where $\hhat{\np}_0$ is the conditional MLE for $\onepoi=0$. Hence, $\hat{\onepoi}_\mathrm{phys} = 0$ for $\hat{\onepoi} < 0$, and $\hat{\onepoi}_\mathrm{phys} = \hat\onepoi$ otherwise. 

For historical reasons, this construction is often described through the lens of using $t = \hat{\onepoi}$ as a test statistic, where it is often referred to as the \textit{Feldman-Cousins construction}~\cite{Feldman:1997qc}. 
Instead of using a fixed ordering rule (a central interval or upper/lower limit), the ordering rule is defined through the LR $t_{\onepoi}^\LR(\data)$ in Eq.~\eqref{eq:tLR_boundary}.
As in the case without boundaries, an acceptance region $t_{\onepoi}^\LR < t_0$ in test-statistic space induces an analogous acceptance region in the space of the $\hat{\onepoi}$ test statistic. 

This is shown in the top panel of Fig.~\ref{fig:Neyman_boundary} for a Gaussian parameter, $\onepoi$, with standard deviation $\sigma_\onepoi=1$. The acceptance region (``Neyman belt''), is obtained by solving Eq.~\eqref{eq:Neyman_belt} and adopting the ordering rule $t_{\onepoi}^\LR < t_0$ for $(1-\alpha) = 68\%$ (red shaded region) and $95\%$ (blue shaded region). 
Without boundaries, a cut on $\chi^2$ implied a central acceptance region in $\onepoi$. In the presence of a boundary and with the modified relations that it implies, the cut on the \emph{physical} profile-LR (vertical black dashed line) also leads to central intervals in $\onepoi$, but with a more complex shape, which gradually evolves towards one-sided intervals as the boundary is approached (horizontal red/blue dashed lines).

Alternatively, one can use the profile LR, $t = t_{\onepoi}^\LR$  (Eq.~\ref{eq:tLR_boundary}) as a test statistic to define acceptance regions for the Neyman construction as illustrated in the center panel of Fig.~\ref{fig:Neyman_boundary}. Analogously to the case without boundaries, for a given C.L. $(1-\alpha)$ one can define an acceptance region as $t_{\onepoi}^\LR < t_0$ (red/blue shaded regions). However, since close to the boundaries, the distribution deviates from a $\chi^2$ distribution (as described below), the Neyman band is not a constant vertical band anymore but rather shrinks towards the boundary.

With the Neyman belt constructed, we can finish the interval construction by considering the test statistic value $t_{\onepoi}^\LR(\data_\obs)$ for the observed data as a function of $\onepoi$ (black dashed line). Two cases must be differentiated: for $\hat{\onepoi}>0$, the familiar parabolic shape is recovered, reaching $t_{\onepoi}^\LR = 0$ at the best-fit value, $\hat\onepoi$, while for $\hat{\onepoi}<0$, the parabola is shifted and the $t_{\onepoi}^\LR = 0$ is reached at the best possible \emph{physical} value, $\hat\onepoi_\mathrm{phys} = 0$. The interval is constructed in both cases in the familiar way by observing, where $t_{\onepoi}^\LR(\data_\obs)$ enters and exits the (now modified) Neyman band (horizontal red/blue dashed lines). It is evident from this construction that this will never lead to empty intervals. Furthermore, the intervals will smoothly evolve from a one-sided interval to a two-sided interval. The described construction can be viewed as a \textit{boundary-corrected} graphical construction, where the cutoff values are not obtained from Wilks' theorem but are the modified ones.

As stated above, the presence of the boundary leads to a deviation from Wald's relation (Eq.~\ref{eq:Walds_relation}). 
This is illustrated in the bottom panel of Fig.~\ref{fig:Neyman_boundary} for a Gaussian parameter, $\onepoi$ with $\mu_\true = 0.25$ and $\sigma_\onepoi=1$. The modified LR test statistic, Eq.~\eqref{eq:tLR_boundary} is consistent with Wald's relation (black dashed line) at $\hat\onepoi>0$ but deviates for $\hat\onepoi <0$. 
The situation can be salvaged into a slightly more general relation for $\onepoi < 0$ by considering that:
\begin{equation}
    \begin{split}
    t_{\onepoi}^\LR (\data) &= - 2 \log \frac{\like_{\data}(\onepoi, \hhat{\vec\nu})}{\like_{\data}(0,\hhat{\np}_0)} \\
    &= -2\left(\log \frac{\like_{\data}(\onepoi, \hhat{\vec\nu})}{\like_{\data}(\hat\onepoi,\hat{\np})} -\log  \frac{\like_{\data}(0, \hhat{\vec\nu}_0)}{\like_{\data}(\hat\onepoi,\hat{\np})}\right).
    \end{split}
\end{equation}
With this, the authors of Ref.~\cite{Cowan:2010js} have derived a new relation between the (possibly unphysical) best-fit value $\hat{\onepoi}$ and the LR that respects a physical boundary: 
\begin{equation}
    \label{eq:mod_Wald}
    t_{\onepoi}^\LR(\data) 
    = \begin{cases}
    \frac{(\hat{\onepoi}-\onepoi)^2}{\sigma_{\hat{\onepoi}}^2} - \frac{\hat{\onepoi}^2}{\sigma_{\hat{\onepoi}}^2} 
    = \frac{\onepoi^2}{\sigma_{\hat{\onepoi}}^2} - \frac{2\onepoi\hat{\onepoi}}{\sigma_{\hat{\onepoi}}^2}\;\mathrm{if}\; \hat{\onepoi} < 0, \vspace*{0.2cm}\\
    \frac{(\hat{\onepoi}(x)-\onepoi)^2}{\sigma_{\hat{\onepoi}}^2}\;\mathrm{otherwise}.
    \end{cases}
\end{equation}
That is, the relationship is linear for $\hat\onepoi < 0$ and parabolic for $\hat{\onepoi} > 0$. Consequently, the acceptance region shrinks towards the boundary~\footnote{These modifications to the graphical method are rarely visualized in this manner. We refer the reader to Kyle Cranmer's Lectures on Statistics, particularly Lecture 3 at \url{https://indi.to/D6dtm}.}.
This modified relationship implies the necessity to adapt Wilks' theorem. Clearly, even in the asymptotic limit, where $\hat{\onepoi}$ approaches a Gaussian distribution, the resulting test statistic distribution (histogram along the $y$ axis in the bottom panel of Fig.~\ref{fig:Neyman_boundary}) is not $\chi^2$-distributed (black dashed line). This deviation from Wilks' theorem is reflected in the shape of the acceptance regions: as more of the linear branch with $\onepoi < 0$ gets populated by mock samples, the limit of the acceptance regions tend to be lower (for $t_\obs = 0.25$, $\sim 0.6$ instead of $1$ at $68\%$ and $\sim 2.9$ instead of $4$ at $95\%$ C.L.). This is reflected in the lower cutoff of the Neyman belt towards $\onepoi=0$ in the center panel of Fig.~\ref{fig:Neyman_boundary}. Hence, the correction from the boundary leads to \textit{tighter} constraints compared to a naive graphical construction that ignores the boundary. 

It is useful to consider two limiting cases here: At the boundary $\onepoi=0$, the modified relationship is a ``half-parabola'': the test statistic vanishes for $\hat{\onepoi} < 0$ and is parabolic on the positive branch. This will lead to a ``half-$\chi^2$'' distribution, where half of the probability mass is concentrated at 0. Far away from the boundary, the modification does not matter since for $\onepoi \gg 0$ the distribution will be standard $\chi^2$-distributed. In between, a significant fraction of events will populate the linear branch in the unphysical region, leading to a mixed distribution.

To summarize, the physical boundaries require some changes to the asymptotic theory. In particular, the standard Wilks' theorem does not hold and neither does the graphical construction of looking at fixed levels of the profile LR to create intervals quickly, i.e. the ``graphical method.'' The relationships can be consistently modified, however, within the assumptions of asymptotic behavior and these modifications naturally lead to a modified graphical construction and the Feldman-Cousins prescription of confidence intervals.

\subsection{Checking for a breakdown of asymptotic behavior}

The constructions above rely on asymptotic behavior. However, for a general model $p(\data|\allpar)$ it is not easy to determine whether the asymptotic relations hold without further inspection. It is, therefore, crucial to check whether the assumptions hold, before proceeding to use e.g.\ the graphical or Feldman-Cousins methods. The matter is complicated by the fact that some aspects of the asymptotic behavior can be reached before others. In particular the following should be checked using mock data samples:

\paragraph{Asymptotic normality of MLE estimators:} 
The distribution of best-fit values $\hat{\poi}$ from a maximum-likelihood optimization should follow a Gaussian distribution. Furthermore, the variance of the $\hat{\poi}$ must also be estimated.

\paragraph{Wald's relation and independence of nuisance parameters:}
For a given model $\data \sim p(\data|\poi,\np)$, the best-fit values $\hat{\poi}$ and the profile LR test statistic $t_{\poi}^\mathrm{LR}$ should follow the parabolic relation from Eq.~\eqref{eq:Walds_relation}. Moreover, this should be independent of the value of $\np$, i.e. the relation should hold even when varying the nuisance parameters.

\paragraph{Wilks' theorem:} 
In cases without a boundary, the familiar $\chi^2$ distribution should hold for the sampling distribution of $t_{\poi}^\mathrm{LR}$. With a boundary, it should deviate from the $\chi^2$ distribution in accordance with the modification discussed in Sec.~\ref{sec:Gaussian_boundaries}.

%%%%%%%%%%%%%%%%%%%%%%%%%%%%%%%%%%%%%%%%%%%%%%%%%%%%%%%%%%%%%%%%%%%%%%

\section{\lowercase{\pinc}: Simulated-annealing minimization interfaced with \texttt{CLASS}}
\label{sec:pinc}

Computing profile likelihoods amounts to minimizing the negative log-likelihood, $-\log\like_{\data_\obs}(\onepoi, \hhat{\np})$, for different fixed values of the parameter of interest $\onepoi$ to obtain the conditional MLEs, $\hhat{\vec\nu}$, as well as the computation of one global MLE $-\log\like_{\data_\obs}(\hat\onepoi, \hat{\np})$ (see Eq.~\ref{eq:t_PLR}). In cosmological applications, the likelihood typically depends on the theory predictions obtained via Boltzmann solvers like \CLASS\ ~\cite{Blas:2011rf} and \texttt{CAMB}~\cite{Lewis:1999bs,Howlett:2012mh}. Depending on the cosmological model, one evaluation of the likelihood via a Boltzmann solver can take up to several seconds. Hence, an efficient minimization algorithm is essential to obtain (conditional) MLEs. 

Many minimizers like \texttt{minuit} \cite{James:1994vla} and \texttt{bobyqa} 
\cite{Powell:2009} are based on the evaluation of gradients. Gradient-based minimizers have been used in cosmological settings in e.g.\ \cite{Planck:2013nga, Henrot-Versille:2016htt}, which requires tuning of the precision settings of Boltzmann solvers like \CLASS\ and \texttt{CAMB}. However, cosmological likelihoods can be noisy due to the use of different approximation schemes in different parts of the parameter space or insufficient precision. Therefore, simulated-annealing-based algorithms often outperform gradient-based methods, see e.g.\ \cite{Hannestad:2000wx, Schoneberg:2021qvd, Reeves:2022aoi}. 

The idea behind simulated annealing is the following: The minimizer walks through the likelihood landscape, (similar to MCMC chains) with step size $F$ and acceptance probability given by
\begin{equation}
    P(\like_i, \like_{i+1}) \sim \exp\left(-\frac{\like_{i+1} - \like_{i}}{T} \right),
\end{equation}
where $\like_i$ is the current position and $\like_{i+1}$ the proposed new position in the likelihood landscape. $T$ is the temperature, which determines how sensitive the algorithm is to differences in the likelihood. Opposed to (Markovian) chains, the step size $F$ and temperature $T$ of the chains change along the way. The chains are initialized with a large step size $F$ and high temperature $T$, which allows them to explore the whole parameter space. $F$ and $T$ are then successively decreased, which makes the chains more sensitive to potential wells in the -log-likelihood surface and will eventually trap them in the (global) minimum. 

With this paper, we make our code \pinc\ (``\textbf{p}rofiles \textbf{in} \textbf{c}osmology) available\footnote{ \url{https://github.com/LauraHerold/pinc}}, which also includes the notebooks to reproduce the figures in this paper. \pinc\ employs the simulated annealing scheme inspired from \cite{Schoneberg:2021qvd}. It interfaces with the MCMC sampler \Montepython~\cite{Audren:2012wb, Brinckmann:2018cvx} and submits chains with decreasing step size $F$ and temperature $T$ using \Montepython's built-in settings of $F$ and $T$. This allows us to keep the code very minimalistic: \pinc\ consists of only three short scripts, which automatically set the relevant parameters in \Montepython, submit the minimization chains, and analyze the results. The analysis assumes a Gaussian distribution but takes into account corrections from physical boundaries (Sec.~\ref{sec:Gaussian_boundaries}). Hence, no installation is necessary as the three \pinc\ scripts can easily be copied in and adapted to any existing \Montepython\ installation. An extension of the framework is left for future work. 

As of now, we are aware of four other public profile likelihood codes interfaced with cosmological codes: \texttt{CAMEL}\footnote{\url{http://camel.in2p3.fr}} \cite{Henrot-Versille:2016htt}, which makes use of the \texttt{minuit} minimizer;
\texttt{PROSPECT}\footnote{\url{https://github.com/AarhusCosmology/prospect_public}} \cite{Holm:2023uwa}, and \texttt{PROCOLI}\footnote{\url{https://github.com/tkarwal/procoli/}} \cite{Karwal:2024qpt}, which are based on simulated-annealing minimization; and 
\texttt{Cobaya-fork}\footnote{\url{https://github.com/ggalloni/cobaya/tree/profile_sampler}}, which uses the built-in Cobaya minimizer.

%%%%%%%%%%%%%%%%%%%%%%%%%%%%%%%%%%%%%%%%%%%%%%%%%%%%%%%%%%%%%%%%%%%%%%

\section{Data sets and mock data generation}
\label{sec:data_and_mocks}

In order to probe the probability distribution of cosmological parameters, we need to generate mock realizations of the data. Here, we want to focus on \Planck\ CMB data since it gives the most competitive constraints on most cosmological parameters. However, the full \Planck\ data set is too complex to be simulated in large numbers \cite{Planck:2019nip}:
the low-$\ell$ likelihoods ($2\leq \ell < 30$, \texttt{Commander}/\texttt{SimAll}) are computed at the level of the pixel map since the power spectrum is non-Gaussian at these scales; 
and the high-$\ell$ likelihood ($30 \leq \ell \leq 2500$ in temperature and $30 \leq \ell \leq 2000$ in polarization, \texttt{Plik}) is based on ``pseudo-$C_\ell$'s'' from different frequency channels, which introduces 47 nuisance parameters to model instrument noise and foregrounds. Since these likelihoods start from data levels more complicated than the cleaned $C_\ell$'s it is non-trivial to generate mock realizations of these data and is beyond the scope of this paper. Therefore, in this paper, we use the simpler \Pliklite\ likelihood, which we describe in the next section.

\subsection{Generating mock \Pliklite\ data}
\label{sec:generating_mocks}

The \Planck\ \Pliklite\ likelihood is a nuisance pre-marginalized version of the \texttt{Plik} likelihood. 
Instead of using the full multi-frequency likelihood with all nuisance parameters, it first extracts CMB temperature and polarization power spectra, while marginalizing over foreground and noise contributions, leaving the \Pliklite\ likelihood with only one nuisance parameter, $A_\mathrm{\textit{Planck}}$, the calibration of the overall amplitude of the power spectra. Hence, the \Pliklite\ likelihood can be written down as a simple Gaussian likelihood\footnote{Note that this does not mean that the likelihood in the cosmological parameters is automatically Gaussian.}:
\begin{equation}
    \label{eq:Plik_lite_like}
    \ln \like(\hat{\vec C}(\data)|\vec{C}(\vec\theta)) = \frac{1}{2} [\hat{\vec C}(\data)- \vec{C}(\vec\theta)]^T \Sigma^{-1}[\hat{\vec C}(\data) - \vec{C}(\vec\theta)],
\end{equation}
where $\hat{\vec C}(\data) = [\hat C^\mathrm{TT}_\ell,\ \hat C^\mathrm{TE}_\ell,\ \hat C^\mathrm{EE}_\ell]$ denotes the temperature and E-mode polarization (TT, TE, EE) CMB-only power spectrum multipoles estimated from the raw data $\data$. In the context of frequentist inference, $\hat{\vec C}(\data)$ plays the role of $\data_\obs$. $\vec{C}(\vec\theta)$ denotes the theory model depending on the (cosmological) parameters $\vec \theta$. $\Sigma$ denotes the covariance matrix published by \Planck, which also contains foreground and noise uncertainty.

For our mock spectra, we assume the 2018 \Planck\ best-fit parameters \cite{Planck:2018vyg} as our fiducial ``true'' cosmology, which we quote in Tab.~\ref{tab:fiducial_cosmo}. Moreover, we assume two massless and one massive neutrino carrying the total mass, $M_\nu = 0.06\,$eV, except for the $\Lambda$CDM+$M_\nu$ model, where we assume three degenerate-mass neutrinos in order to facilitate direct comparison with the \Planck\ 2018 results \cite{Planck:2018vyg}. We compute the CMB power spectra using the Boltzmann code \CLASS\ \cite{Blas:2011rf}\footnote{\url{http://class-code.net}}  and model non-linear corrections with \texttt{halofit} \cite{Smith:2002dz}. 

We generate mock TT, TE, and EE spectra by drawing spectra from a multivariate Gaussian with mean $C_\ell^\mathrm{(fiducial)}$ and covariance matrix $\Sigma$, where $\Sigma$ is taken from the \Pliklite\ likelihood, Eq.~\eqref{eq:Plik_lite_like}. We bin the spectra with the scheme described in App.~\ref{sec:Pliklite}. The resulting mock spectra can then easily be inserted in the public \Planck\ \texttt{clik} likelihood.\footnote{\url{https://github.com/benabed/clik}} 

Since the \Pliklite\ likelihood contains only scales $\ell \geq 30$, it is only sensitive to the combination $A_s e^{-2\tau_{\mathrm{reio}}}$, where the degeneracy between the optical depth to reionization, $\tau_\reio$, and the amplitude of the primordial power spectrum, $A_s$, is usually broken by the low-$\ell$ temperature and polarization data. Therefore, the only nuisance parameter of the \Pliklite\ likelihood, $A_{Planck}$, is fully degenerate with $A_s$.  Thus -- if not otherwise indicated -- we fix $\tau_\mathrm{reio} = 0.0543$ and $A_{Planck} = 1$ to their fiducial values, which for $\tau_\mathrm{reio}$ corresponds to its best-fit value for full \Planck\ 2018 data \cite{Planck:2018vyg}. 
We check that the posteriors of the cosmological parameters of the \Pliklite\ likelihood in this setup are consistent with the full \Planck\ data for the three cosmological models we consider in this paper ($\Lambda$CDM, $\Lambda$CDM+$M_\nu$, $w_0$CDM) in App.~\ref{sec:posteriors}.
\begin{table}
    \caption{\Planck\ 2018 best-fit parameters used as the fiducial cosmology to generate mock \Planck-lite data. Moreover, we fix the only nuisance parameter to the theoretically expected value, $A_{Planck}=1$.}
    \centering
    \begin{tabular}{|l|c|}
    \hline
    Parameter             & Fiducial value \\ \hline
    $\omega_{b }$         & $0.022383$ \\
    $\omega_\mathrm{cdm}$ & $0.12011$ \\
    $\ln(10^{10}A_{s})$   & $3.0448$  \\
    $n_{s}$               & $0.96605$ \\
    $\tau_\mathrm{reio}$  & $0.0543$  \\
    $h$                   & $0.6732$  \\
%    $A_{Planck}$          & $1$  \\
    \hline
    $M_\nu$               & $0.06\,$eV  \\
    $w_0$                 & $-1$\\
    \hline
    \end{tabular}
    \label{tab:fiducial_cosmo}
\end{table}

\subsection{Methodology and data sets}

For the frequentist and Bayesian analysis in Sec.~\ref{sec:profiles}, we make use of the public MCMC sampler \texttt{MontePython} \cite{Brinckmann:2018cvx} interfaced with the Boltzmann solver \texttt{CLASS} \cite{Blas:2011rf} and use \texttt{GetDist} \cite{Lewis:2019xzd} to visualize posteriors. We
consider the \Pliklite\ likelihood \cite{Planck:2019nip}, referred to as \Planck-lite, the \textit{Planck} TT, TE, EE and lensing data \cite{Planck:2018vyg}, referred to as \Planck, as well as BAO data from the 6dF Galaxy Survey \cite{Beutler:2011hx}, from the Sloan Digital Sky Survey (SDSS) DR 7 Main Galaxy Sample \cite{Ross:2014qpa} and from the SDSS Baryon Oscillation Spectroscopic Survey (BOSS) \cite{BOSS:2016wmc}, referred to as BAO.

%%%%%%%%%%%%%%%%%%%%%%%%%%%%%%%%%%%%%%%%%%%%%%%%%%%%%%%%%%%%%%%%%%%%%%

\section{Results: Distribution of mock \textit{Planck} data}
\label{sec:results}

Since it is difficult to verify whether the asymptotic limit or Wilks' theorem holds in practice, in this section, we explicitly probe the distribution of the LR test statistic under mock \Planck-lite spectra in order to verify whether it is consistent with the predictions by Wilks and Wald (Sec.~\ref{sec:wilks}). 
Note, however, that we conduct this check only for one set of parameters called the fiducial cosmology. To verify that the asymptotic limit is reached, it is necessary to consider many different cosmologies. Hence, we can only assess whether the asymptotic assumption does \textit{not} hold if the check for the fiducial cosmology fails. Nevertheless, we get an indication of asymptoticity if the mocks follow the predictions by Wilks and Wald for the fiducial cosmology.

We consider three different cosmological models: the $\Lambda$CDM model; a $\Lambda$CDM+$M_\nu$ model with the total neutrino mass, $M_\nu$, as a free parameter; and a $w_0$CDM model with the equation of state of dark energy, $w_0$, as a free parameter. 
To probe the distribution of the LR test statistic, we generate mock \Planck-lite data, $\data_\mock$, as described in Sec.~\ref{sec:data_and_mocks} assuming the \Planck\ 2018 best fit as our fiducial cosmology (Tab.~\ref{tab:fiducial_cosmo}). For each of the mock CMB spectra, we compute the following profile LR using \pinc\ (c.f.\ Eq.~\ref{eq:t_PLR}):
\begin{equation}
    \label{eq:LR_mocks}
    t_{\onepoi_\true}^\LR(\data_\mock) 
    = - 2 \log\left(\frac{\like_{\vec \data_\mathrm{mock}}(\onepoi_\true, \hhat{\np})}{\like_{\vec \data_\mathrm{mock}}(\hat\onepoi, \hat{\np})} \right),
\end{equation}
where the parameter of interest, $\onepoi$, is one of the cosmological parameters. As discussed in Sec.~\ref{sec:Pliklite}, we fix $\tau_\mathrm{reio}$ and  $A_{Planck}$ as indicated in Tab.~\ref{tab:fiducial_cosmo}, which leaves us with five cosmological parameters: the dimensionless Hubble constant $h= H_0/(100\, \mathrm{km/s/Mpc})$, the cold dark matter energy fraction $\omega_\cdm = \Omega_\cdm h^2$, the baryon energy fraction $\omega_b = \Omega_b h^2$, the amplitude $A_s$ and spectral index $n_s$ of the primordial power spectrum. We evaluate the numerator for the parameter of interest fixed to the fiducial cosmology, $\onepoi = \onepoi_\true$. 

We conduct two types of checks: checks with fixed nuisance parameters, where $\np$ is empty; and checks with varying nuisance parameters, where the nuisance parameters are profiled over.\footnote{Note that we are using the term ``nuisance parameters'' for all \textit{cosmological} parameters except the parameter of interest. This is different from the conventional usage of this word, where nuisance parameters describe technical non-cosmological parameters.} For each of the checks, we specify explicitly the respective $\allpar = (\onepoi,\np)$ that is assumed.

\subsection{Wilks \& Wald in \LCDM}
\label{sec:checks_LCDM}

\begin{figure}
    \centering
    \vspace*{-0.3cm} 
    \subfigure{\includegraphics[scale=0.5]{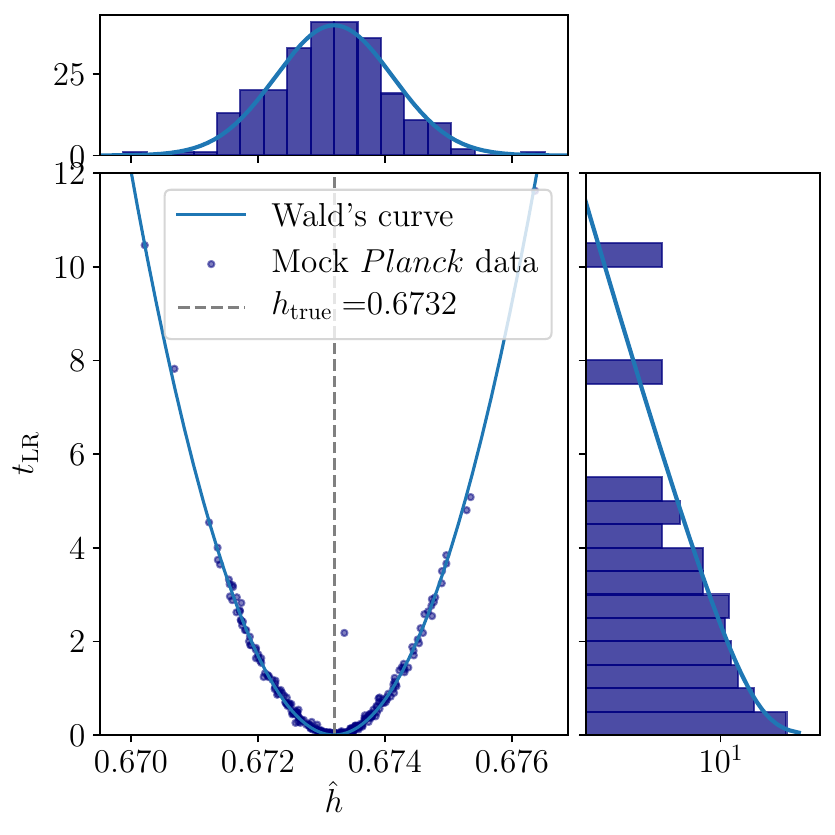}}
    \includegraphics[scale=0.5]{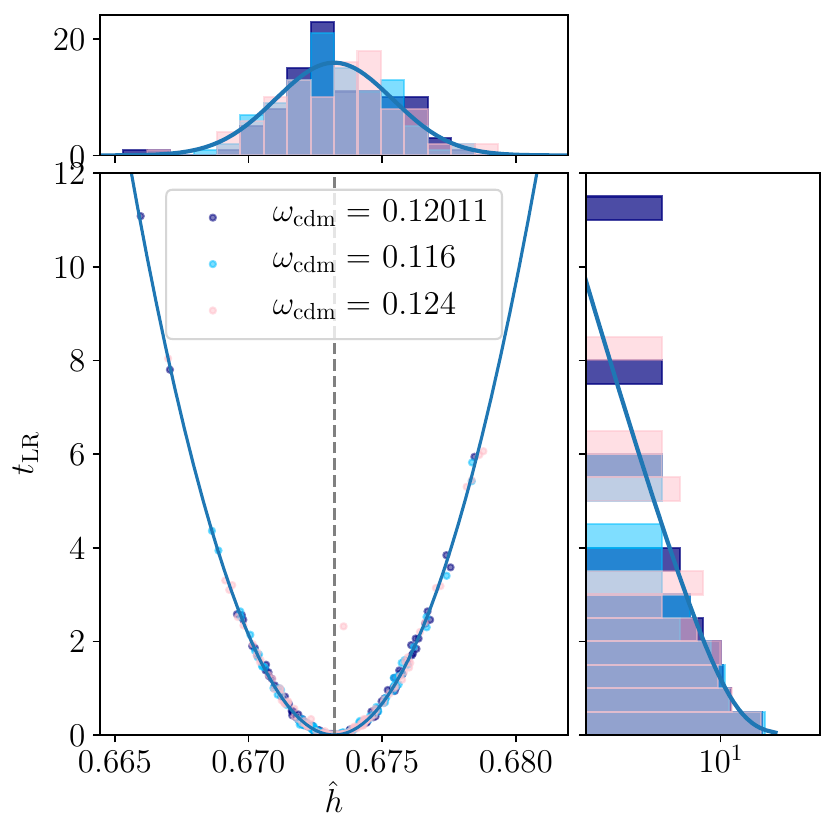}
    \includegraphics[scale=0.5]{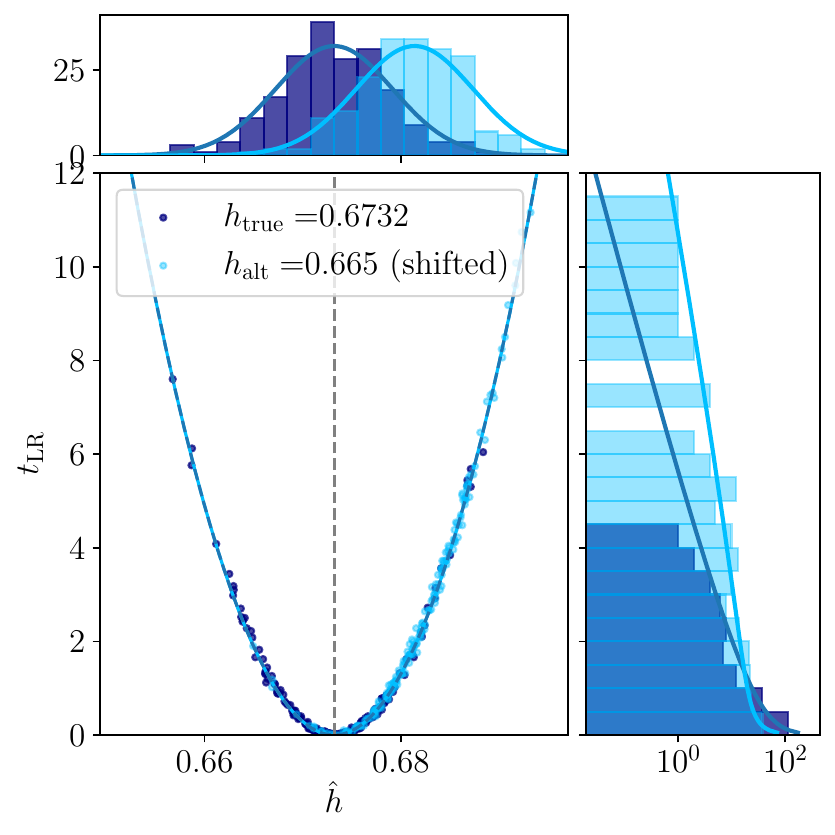}
    \vspace*{-0.3cm} 
    \caption{$t_\mathrm{LR}$,  Eq.~\eqref{eq:LR_mocks}, as a function of the inferred MLE of the dimensionless Hubble parameter $\hat\onepoi=\hat h$ for \Planck\ mock data within \LCDM. \textit{Top: } all parameters except $h$ are fixed. \textit{Center: } we treat $\omega_\cdm$ as a free nuisance parameter and create mocks with three different values of $\omega_\cdm$. \textit{Bottom:} we treat all remaining \LCDM\ parameters as free nuisance parameters and consider the alternative hypothesis $h_\mathrm{alt} = 0.665$. All checks indicate that Wilks' theorem holds in \LCDM\ (see text).}
    \label{fig:LR_hist_LCDM}
\end{figure}

For the $\Lambda$CDM model, we conduct four checks: (\textit{a}) all nuisance parameters fixed, (\textit{b}) only one varying nuisance parameter, (\textit{c}) varying all nuisance parameters, and (\textit{d}) considering an alternative hypothesis. 

\paragraph{Fixed nuisance parameters:} For our first check, we fix all \LCDM\ parameters, except the parameter of interest, $\onepoi$, to the fiducial cosmology, leaving us with a one-dimensional likelihood surface. As the parameter of interest, we choose the dimensionless Hubble parameter $\onepoi = h$ here, but show the results for the other five $\Lambda$CDM parameters in App.~\ref{sec:LR_hist_LCDM_fixed_NP}. We generate 250 mock CMB spectra, $\data_\mock$, with the fiducial cosmology and compute Eq.~\eqref{eq:LR_mocks} for
\begin{equation}
    \onepoi = h,\quad \np = \{\},
\end{equation}
where all other parameters except $h$ are held fixed and thus $\np$ is empty. In practice, for each $\data_\mock$ the numerator of Eq.~\eqref{eq:LR_mocks} requires one evaluation of the likelihood for the parameter $\onepoi = h_\true$, with $h_\true$ fixed to the fiducial value, and the denominator requires one minimization with one free parameter to obtain the MLE $\hat\onepoi = \hat h$.

The results of this check are presented in the top panel of Fig.~\ref{fig:LR_hist_LCDM}. 
As described in Sec.~\ref{sec:wilks}, the solid blue line describes ``Wald's curve,'' which is assumed in the asymptotic limit. We describe how we compute Wald's curve with the Asimov data set for mock CMB data in App.~\ref{sec:Asimov}. The blue markers show the value of the LR test statistic $t^\LR$ of the mock spectra in Eq.~\eqref{eq:LR_mocks} as a function of the \textit{inferred} MLE $\hat h$. Note that all mock CMB spectra have been generated with the same fiducial cosmology but due to the statistical fluctuations of the spectra around the fiducial cosmology, the MLE can differ from the true value of the parameter (as indicated by the vertical dashed line). 
We observe that $t^\LR$ of the mock CMB spectra follow closely Wald's curve. 

Moreover, the histograms along the $x$ axis (top subplot) show that the distribution of the MLE $\hat\onepoi$ of the mock CMB spectra is consistent with a Gaussian distribution normalized to the number of mocks as indicated by the blue solid line. 
The histograms along the $y$ axis (right subplots) show that the distribution of $t^\LR$ of the mock \Planck\ spectra is consistent with a $\chi^2$ distribution with one degree of freedom normalized to the number of mocks as indicated by the blue solid line. 

We show the same check for the other five $\Lambda$CDM parameters $\{\omega_\cdm, \omega_b, A_s, n_s, \tau_\reio \}$ in Fig.~\ref{fig:LR_hist_LCDM_fixed_NP} in App.~\ref{sec:LR_hist_LCDM_fixed_NP}. For all $\Lambda$CDM parameters, we find good agreement with Wald's curve apart from some outliers which are most likely due to failed minimizations.\footnote{The failed minimizations are possibly because of chains getting stuck in local minima, which is common, especially for cases with many free parameters.}

Hence, this first test indicates that in the simple case with all nuisance parameters fixed, the distribution of the mock \Planck-lite spectra follows closely the predicted Wald's curve. The $\chi^2$ distribution is a good description of the distribution of $t^\LR$ when all nuisance parameters are fixed, as predicted by Wilks' theorem. 

\paragraph{One varying nuisance parameter:} Next, we explore the dependence of the distribution on the value of one particular nuisance parameter. Since $h$ has the strongest degeneracy with the cold dark matter fraction $\omega_\cdm$, we compute Eq.~\eqref{eq:LR_mocks} for 
\begin{equation}
    \begin{split}
    &\onepoi = h,\quad \np = \omega_\cdm,\\
    &\data_\mock = \{\data_\mock^{\omega_{\cdm,\true}}\}_{\omega_{\cdm,\true} = \{ 0.116,\ 0.121011,\ 0.124 \}},
    \end{split}
\end{equation}
where the mock CMB spectra, $\data_{\mock}^{\omega_{\cdm, \true}}$, are generated with three different true values of $\omega_\cdm$, respectively. The second value, $\omega_{\cdm,\true} = 0.121011$, corresponds to the fiducial value of $\omega_\cdm$. 
All other parameters are held fixed to their fiducial values (Tab.~\ref{tab:fiducial_cosmo}). With this setup, we generate three sets of 100 mock spectra for each of the three values $\omega_{\cdm,\true}$. 
In practice, the computation of the numerator of Eq.~\eqref{eq:LR_mocks} requires one minimization with one free parameter, $\np = \omega_\cdm$, for each mock spectrum $\data_{\mock}^{\omega_{\cdm, \true}}$, while the denominator requires one minimization with two free parameters, $(h, \omega_\cdm)$, per $\data_{\mock}^{\omega_{\cdm, \true}}$. 

We show the results of this test in the center panel of Fig~\ref{fig:LR_hist_LCDM}, where the different colors correspond to the three different values of $\omega_\cdm$. The test statistic $t^\LR$ as a function of the MLE $\hat h$ follows closely the predicted Wald's curve with no dependence on the value of $\omega_{\cdm,\true}$. The histogram of the mock spectra in bins of $\hat h$ is consistent with a Gaussian distribution and the histogram of the mock spectra in bins of $t^\LR$ is consistent with a $\chi^2$ distribution, regardless of $\omega_{\cdm,\true}$. 

This test indicates that in the simplified case of only one varying nuisance parameter, $\omega_\cdm$, the distribution of the mock CMB spectra for the parameter $h$ is consistent with the predictions by Wilks and Wald. We found that this observation is independent of the value of $\omega_{\cdm,\true}$ used to generate the mock spectra for all three choices of $\omega_{\cdm,\true}$ in this test, which is a key ingredient in the asymptotic theory (Sec.~\ref{sec:wilks}). 

\paragraph{Varying nuisance parameters:} In our next test, we generate 200 mock CMB spectra, $\data_\mock$, with the fiducial cosmology and compute Eq.~\eqref{eq:LR_mocks} for 
\begin{equation}
    \onepoi = h,\quad \np = \{\omega_\cdm, \omega_b, A_s, n_s \}.  
\end{equation}
Hence, the computation of the numerator of Eq.~\eqref{eq:LR_mocks} requires one minimization with five free parameters, $\np$, for each $\vec x_\mock$, while the denominator requires one minimization with six free parameters, $(h, \np)$, per $\data_\mock$. 

We show the results of this check in dark blue color in the bottom panel of Fig.~\ref{fig:LR_hist_LCDM}. Since we have five or six free parameters in the minimizations, respectively, $t^\LR$ and $\hat h$ show a larger scatter. Nevertheless, the distribution of the $t^\LR$ as a function of the MLE $\hat h$ is consistent with Wald's curve. The distribution of the number of mock spectra in bins of $\hat h$ is well described by a Gaussian, and the distribution of the number of mock spectra in bins of $t^\LR$ is well described by a $\chi^2$ distribution. 

\paragraph{Varying nuisance parameters under an alternative hypothesis:} Finally, we want to explore the distribution under an alternative hypothesis and verify that it is distributed as a non-central $\chi^2$ distribution (see Sec.~\ref{sec:wilks}). For that, we use the same mocks as in the previous test but we assume the alternative hypothesis $h_\mathrm{alt} = 0.665$, which differs from the null hypothesis $h_\true = 0.6732$ used to generate the mock spectra $\data_\mock$. Hence, we compute Eq.~\eqref{eq:LR_mocks} for: 
\begin{equation}
    \begin{split}
    &\onepoi = h,\quad \np = \{\omega_\cdm, \omega_b, A_s, n_s \}, \\
    &\onepoi_\true = h_\mathrm{alt} = 0.665.
    \end{split}
\end{equation}
We show the results of this test in light blue color in the bottom panel of Fig.~\ref{fig:LR_hist_LCDM}. The light blue markers have been shifted by $h_\true - h_\mathrm{alt} = 0.0082$ to lie on the same curve as the null hypothesis. We find that also under the alternative hypothesis, the mock spectra follow Wald's curve and the distribution of $\hat\onepoi$ (histogram along the $x$ axis) is consistent with a Gaussian distribution. As expected for the alternative hypothesis, the distribution of $t^\LR$ (histogram along the $y$ axis) is consistent with a \textit{non-central} $\chi^2$ distribution as indicated by the light blue line. 

So, also in our most general test for the \LCDM\ model with all cosmological parameters varying, we find that the distribution of mocks is consistent with Wilks and Wald, both for the null and an alternative hypothesis. This is a good indication that the asymptotic limit and Wilks' theorem hold and the graphical profile likelihood construction will yield correct coverage in the case of the $\Lambda$CDM parameters. Note that this is of course no proof that Wilks' theorem holds for any cosmology other than the fiducial cosmology assumed here.

\subsection{Wilks \& Wald in \LCDM\ $+\ \boldsymbol{M_\nu}$}
\label{sec:checks_Mnu}

\begin{figure}
    \centering
    \includegraphics[scale=0.5]{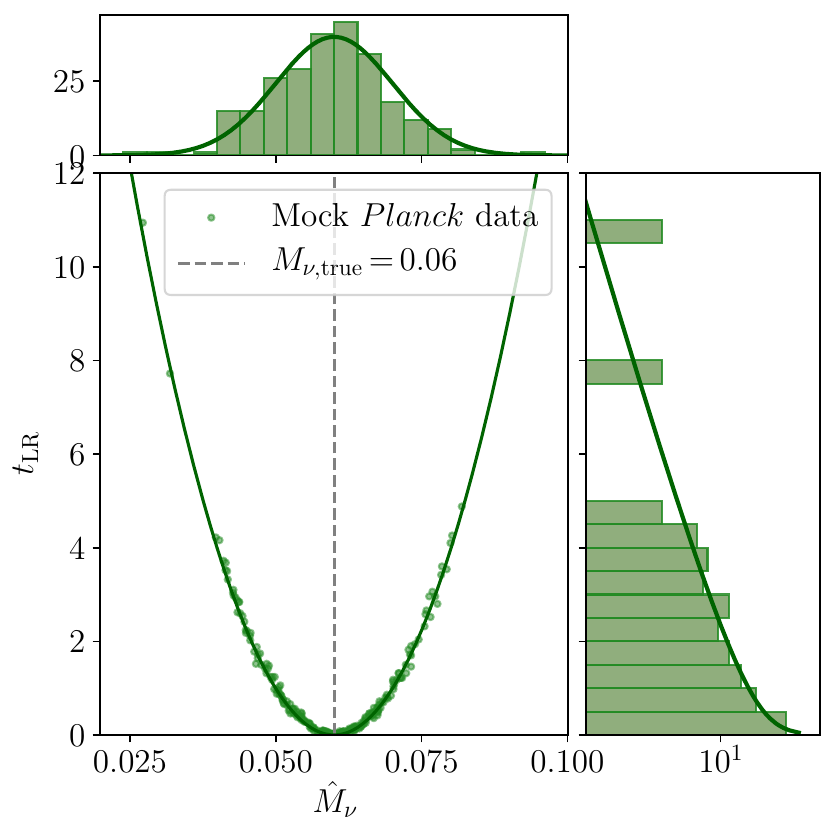}
    \includegraphics[scale=0.5]{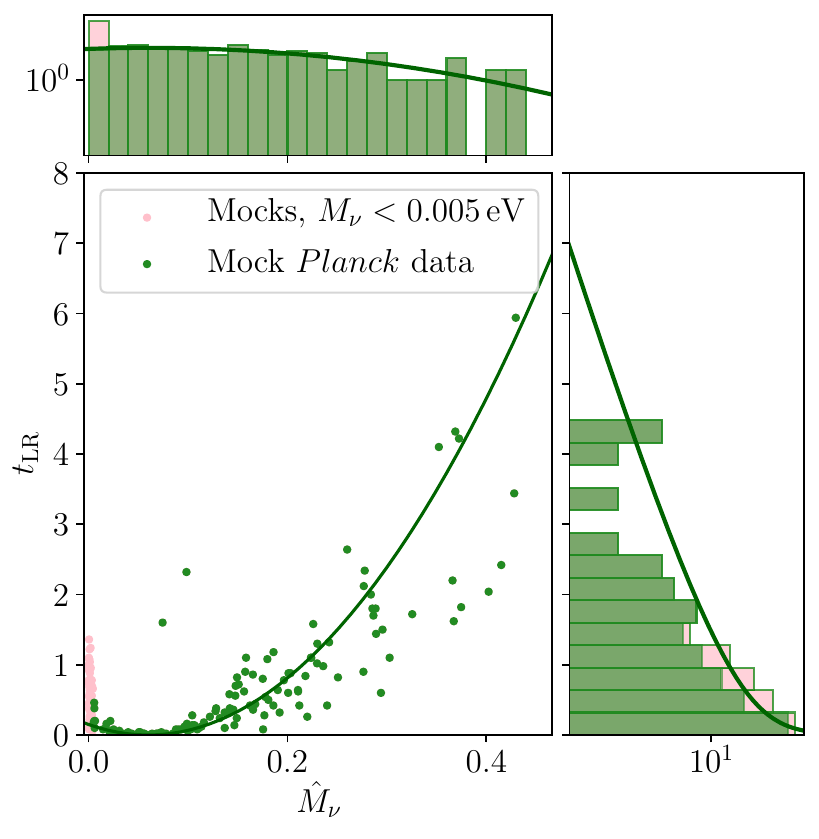}
    \includegraphics[scale=0.5]{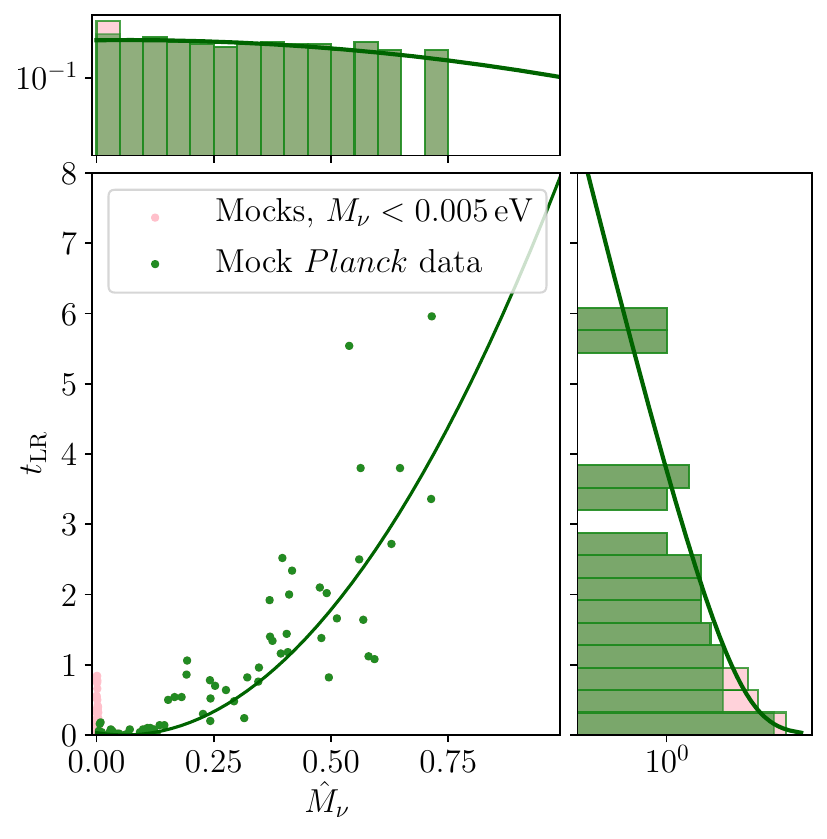}
    \caption{$t_\mathrm{LR}$,  Eq.~\eqref{eq:LR_mocks}, as a function of the inferred MLE of the sum of neutrino masses $\hat\onepoi=\hat M_\nu$ for mock realizations of \Planck\ data within \LCDM+$M_\nu$. \textit{Top: } all parameters except $M_\nu$ are fixed. \textit{Center: } we treat $h$ as a free nuisance parameter. \textit{Bottom:} we treat all \LCDM\ parameters as free nuisance parameters. The checks indicate that $M_\nu$ is distributed as a Gaussian near a boundary (see text).}
    \label{fig:LR_hist_Mnu}
\end{figure}

In this section, we want to explore the distribution of the mock CMB spectra when including the sum of neutrino masses, $M_\nu$, as a free parameter. Since neutrinos are known to have mass, with a minimum sum of $M_\nu > 0.06\,$eV from neutrino oscillation experiments, e.g.~\cite{Esteban:2020cvm}, it is a natural extension of the $\Lambda$CDM model. For the following tests, we generate 250 mock CMB spectra, $\data_\mock$, with $M_{\nu,\true} = 0.06\,$eV. We describe how we compute Wald's curve with the Asimov data set in App.~\ref{sec:Asimov}.

\paragraph{Fixed nuisance parameters:} We first compute Eq.~\eqref{eq:LR_mocks} for: 
\begin{equation}
    \onepoi = M_\nu,\quad \np = \{ \},
\end{equation}
while holding all other parameters fixed to their fiducial value. The results of this check for 250 mocks are presented in the top panel of Fig.~\ref{fig:LR_hist_Mnu}. As for the parameters of the \LCDM\ model, we find good agreement with the predictions by Wilks and Wald. 

\paragraph{One varying nuisance parameters:} For the next check, we compute $t^\LR$ for 200 mock spectra with $\onepoi=M_\nu$ and one nuisance parameter $\np$. We choose $\np = h$ since $h$ has the strongest degeneracy with $M_\nu$ as can be seen from the posterior in Fig.~\ref{fig:posterior_Mnu} in App.~\ref{sec:posteriors}. Hence, we compute Eq.~\eqref{eq:LR_mocks} for: 
\begin{equation}
    \onepoi = M_\nu,\quad \np = h.
\end{equation}
We show the results of this check in the center panel of Fig.~\ref{fig:LR_hist_Mnu}. This is the first check where we find a significant deviation from the curves by Wilks and Wald: since neutrino masses cannot be negative there is a physical border, $M_{\nu} \geq 0$, which leads to an accumulation of points near $M_{\nu} = 0$.
We illustrate this in the center panel of Fig.~\ref{fig:LR_hist_Mnu} by marking all mocks with $\hat M_\nu < 0.005\,$eV with a pink color. This leads to a deviation of the distribution of $\hat M_\nu$ from a Gaussian distribution as can be seen in the histogram along the $x$ axis, whereas the distribution of $t^\LR$ along the $y$ axis still follows a $\chi^2$ distribution. 
Note that this deviation from Wald's curve was not present in the previous check (\textit{a}), since the standard deviation of the distribution in the case with all parameters fixed is too small to approach $M_\nu =0$.
Moreover, we observe an enhanced scatter of the mocks around Wald's curve, which can be partially explained by enhanced noise in the minimization due to the degeneracy between $M_\nu$ and $h$, but might also be indicative of a more fundamental deviation of the distribution of $\hat M_\nu$ from a Gaussian.\footnote{Since the computation of $t^\LR$ is numerically expensive, we leave the computation of $t^\LR$ for more mocks for future work, which would be facilitated by acceleration of the likelihood evaluation, e.g.\ by using an emulator instead of \texttt{CLASS}.}

Hence, in the case of $M_\nu$ with one free nuisance parameter, $h$, the simple graphical profile likelihood construction described in Sec.~\ref{sec:Gauss_PL} would lead to a confidence interval with \textit{incorrect} coverage. However, the distribution we observe appears -- apart from enhanced noise -- consistent with a Gaussian at a physical border in zero, which indicates that the \textit{boundary-corrected} graphical construction (Feldman-Cousins construction) described in Sec.~\ref{sec:Gaussian_boundaries} gives correct coverage. 

\paragraph{Varying nuisance parameters:} We repeat the above exercise for 100 mock spectra but this time we let all $\Lambda$CDM parameters vary, i.e.\ we compute Eq.~\eqref{eq:LR_mocks} for: 
\begin{equation}
    \onepoi = M_\nu,\quad \np = \{h, \omega_\cdm, \omega_b, A_s, n_s \}.
\end{equation}
The results of this check are shown in the bottom panel of Fig.~\ref{fig:LR_hist_Mnu}. As in the case with only $h$ as a free parameter, we find a large accumulation of points in $M_\nu = 0$, which leads to a deviation from the Gaussian distribution of $\hat M_\nu$. Moreover, as previously we find a larger scatter of the mocks around Wald's curve. 
Hence, as discussed in the previous test (\textit{b}), the simple graphical profile likelihood construction leads to a confidence interval with incorrect coverage and the boundary-corrected graphical construction needs to be used.

\subsection{Wilks \& Wald in $\boldsymbol{w_0}$CDM}
\label{sec:checks_w0}

\begin{figure}
    \centering
    \includegraphics[scale=0.5]{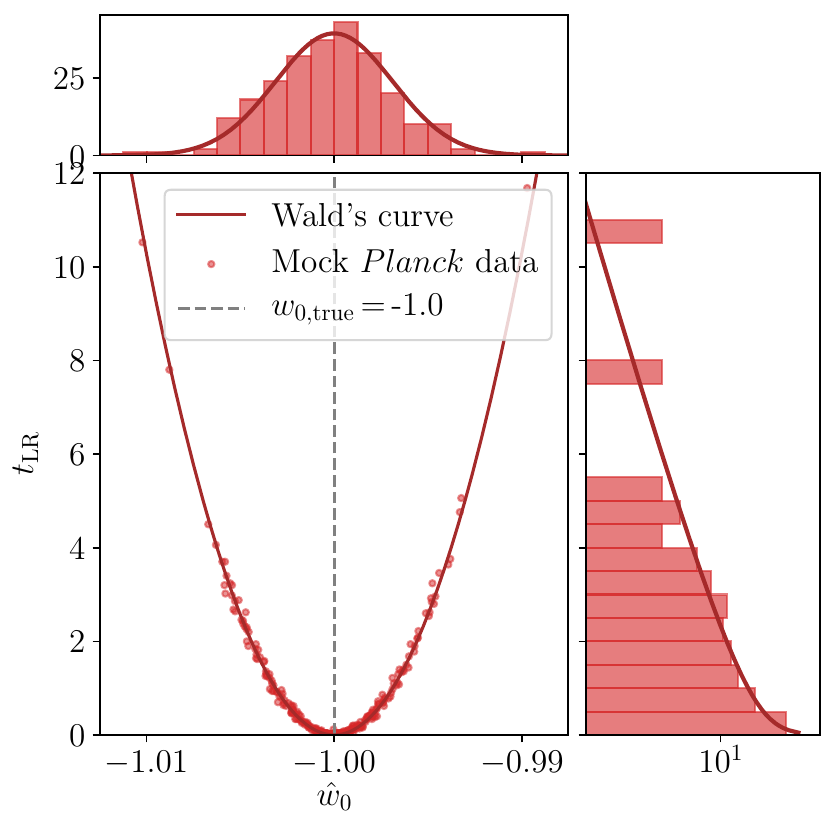}
    \includegraphics[scale=0.5]{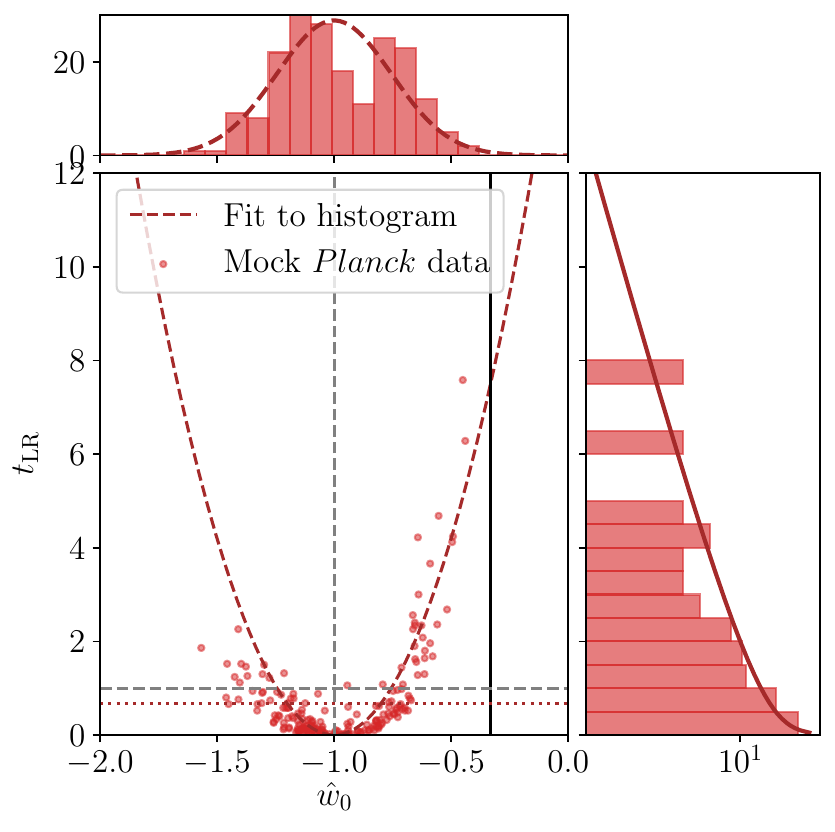}
    \caption{$t_\mathrm{LR}$,  Eq.~\eqref{eq:LR_mocks}, as a function of the inferred MLE of the equation of state of DE $\hat\onepoi=\hat w_0$ for mock realizations of \Planck\ data within $w_0$CDM. \textit{Top: } all parameters except $w_0$ are fixed. \textit{Bottom:} we treat all \LCDM\ parameters as free nuisance parameters. The checks indicate that $w_0$ does \textit{not} follow Wald's curve (see text).}
    \label{fig:LR_hist_w0}
\end{figure}

As the third cosmological model, we consider a $w_0$CDM model with a dark energy component with the equation of state
\begin{equation}
    \label{eq:EOS_w0}
    w_0 = \frac{p_\mathrm{DE}}{\rho_\mathrm{DE}},
\end{equation}
as a free parameter. The mocks in our fiducial cosmology are generated with a cosmological constant, i.e.\ $w_0=-1$. For this model, we conduct two checks. 

\paragraph{Fixed nuisance parameters:} First, we fix all \LCDM\ parameters and compute Eq.~\eqref{eq:LR_mocks} for 250 mock spectra with: 
\begin{equation}
    \onepoi = w_0,\quad \np = \{ \}.
\end{equation}
The top panel in Fig.~\ref{fig:LR_hist_w0} shows the results of this check. As in the other two models, we find good agreement with the predictions by Wilks and Wald when all nuisance parameters are fixed. 

\paragraph{Varying nuisance parameters:} In this final check, we compute Eq.~\eqref{eq:LR_mocks} for 200 mock spectra with: 
\begin{equation}
    \onepoi = w_0,\quad \np = \{h, \omega_\cdm, \omega_b, A_s, n_s \}.
\end{equation}
The analysis of this model is complicated by several factors. The parameter range is restricted to $w_0<-\frac{1}{3}$ since this corresponds to the regime of accelerated expansion. To avoid the minimizations exploring unphysical regimes, we further restrict the parameter range of $h \in [0.2,\ 1.2]$, which leads to an effective restriction of roughly $w_0 \gtrsim -2$ (c.f.\ Fig~\ref{fig:posterior_w0}). 

The results of this check are shown in the bottom panel of Fig.~\ref{fig:LR_hist_w0}.
We find evidence of a deviation from Wald's curve: the CMB mocks (red markers) are not fit by a parabola and show an asymmetry of the two arms of the parabola for $\hat w_0 < -1$ compared $\hat w_0 > -1$. Moreover, the histogram along the $x$ axis (top subplot) shows a bimodal distribution with fewer mocks than expected at $w_0=-1$. To compute Wald's curve, we attempted to compute the standard deviation from the Asimov data as described in App.~\ref{sec:Asimov}, however, the profile in $w_0$ of the Asimov data set represents an asymmetric function not resembling a parabola (see bottom panel of Fig.~\ref{fig:PLs_Asimov}). 

This asymmetry can be explained by the weak sensitivity of \Planck(-lite) data to very negative $w_0$, which is due to the almost perfect degeneracy between $w_0$ and $h$, which can be broken by including BAO data (see the posterior of $w_0$CDM in Fig.~\ref{fig:posterior_w0}).
Moreover, note that different physical models apply (and are used in \texttt{CLASS} \cite{Blas:2011rf}) in the \textit{phantom} regime, $w_0<-1$, and in the \textit{quintessence} regime, $-1<w_0<-\frac{1}{3}$. Further, the restriction of the range $-2\lesssim w_0 < -\frac{1}{3}$ can contribute to the observed deviation from Gaussianity. These factors could explain the asymmetric distribution of $w_0$. 

Hence, instead of computing Wald's curve from the Asimov data set, we obtain the standard deviation $\sigma_\mathrm{hist}$ from a fit to the histogram of $\hat w_0$ (top subplot in the bottom panel of Fig.~\ref{fig:LR_hist_w0}). We show the parabola $(\hat w_0 - w_{0,\true})^2/\sigma_\mathrm{hist}^2$ as the red-dashed line in Fig.~\ref{fig:LR_hist_w0}). We find that the mocks do not lie on the such constructed parabola but are distributed around it in an asymmetric way. This corroborates that the likelihood of $w_0$ is not well described by a Gaussian. 

Moreover, we empirically obtain the boundary of the $68\%$ acceptance region in $t^\LR$ such that 68 of the 100 mocks lie below this cutoff. We find that $68\%$ of the mocks lie below $t_\mathrm{emp}^{68\%} = 0.68$ (horizontal red dotted line in bottom panel of Fig.~\ref{fig:LR_hist_w0}). This is lower than the expected value $t_\mathrm{Gauss}^{68\%} = 1$ for a Gaussian (horizontal grey dashed line in the bottom panel of Fig.~\ref{fig:LR_hist_w0}). Even though the total number of mocks is low due to the high numerical cost of the minimization, this is an indication that $t^\LR$ does not follow a $\chi^2$ distribution. 

Together, these points indicate that the asymptotic regime does not apply for $w_0$ with varying nuisance parameters and that the graphical profile likelihood construction will give incorrect coverage, which in this case \textit{cannot} be remedied by the boundary-corrected/Feldman-Cousins construction. This model warrants further investigation with the use of more mock realizations, which is beyond the scope of this paper due to the prohibitive computational cost. Hence, in this case, we suggest using a full Neyman construction to obtain reliable intervals, which will only be feasible with a considerable speed-up of the likelihood evaluation, e.g.\ by the use of emulators \cite{Auld:2006pm, SpurioMancini:2021ppk, Arico:2021izc, Gunther:2022pto, Nygaard:2022wri}.

%%%%%%%%%%%%%%%%%%%%%%%%%%%%%%%%%%%%%%%%%%%%%%%%%%%%%%%%%%%%%%%%%%%%%%

\section{Constraints on cosmological parameters using the profile likelihood}
\label{sec:profiles}

In this section, we apply the graphical profile likelihood construction to constrain the parameters of the $\Lambda$CDM, $\Lambda$CDM+$M_\nu$ and $w_0$CDM models, and compare the results with Bayesian credible intervals. We compute profile likelihoods for three different data sets: the \Planck-lite pre-marginalised TT, TE, EE likelihood~\cite{Planck:2019nip}, the full \Planck\ TT, TE, EE and lensing data \cite{Planck:2018vyg}, and full \Planck\ data combined with BAO data from 6dF, SDSS, and BOSS \cite{Beutler:2011hx,Ross:2014qpa,BOSS:2016wmc}. We summarize our results in Tab.~\ref{tab:constraints}.

\begin{table*}[]
    \centering
    \begin{tabular}{|c|cc|cc|cc|}
    \hline
    &\multicolumn{2}{c|}{$H_0$ $\big[\frac{\mathrm{km}}{\mathrm{s}\, \mathrm{Mpc}}\big]$ ($\Lambda$CDM)}                     & \multicolumn{2}{c|}{$M_\nu$ [eV]}         & \multicolumn{2}{c|}{$w_0$} \\ 
    \hline
    Data set     &Frequentist         &Bayesian                  &Frequentist &Bayesian &Frequentist          &Bayesian                  \\
    \hline
    \Planck-lite &$67.03 \pm 0.56$ &$66.99_{-0.63}^{+0.61}$ &$<0.16$ &$<0.29$ &($-2.37 \pm 0.83$)   & $<-0.83$ \\
    \Planck      &$67.42 \pm 0.54$ &$67.37_{-0.56}^{+0.53}$ &$<0.18$ &$<0.25$ &($-2.12 \pm 0.58$)   &$<-1.03$   \\
    \Planck+BAO  &$67.68 \pm 0.42$ &$67.71_{-0.43}^{+0.42}$ &$<0.11$ &$<0.12$ &($-1.044 \pm 0.052$) &$-1.040_{-0.057}^{+0.060}$\\
    \hline
    \end{tabular}
    \caption{Frequentist and Bayesian constraints of $H_0 = 100\, h\, \frac{\mathrm{km}}{\mathrm{s}\, \mathrm{Mpc}}$, $M_\nu$ and $w_0$ under CMB (\Planck-lite, \Planck) and BAO (6dF, SDSS, BOSS) data. The frequentist constraints on $w_0$ (in parentheses) might not have correct coverage and are approximate. We quote central intervals at $68\%$ C.L.\ and upper limits at $95\%$ C.L.}
    \label{tab:constraints}
\end{table*}

\subsection{Profiles in $h$ ($\Lambda$CDM)}

\begin{figure}
    \centering
    \includegraphics[scale=0.64]{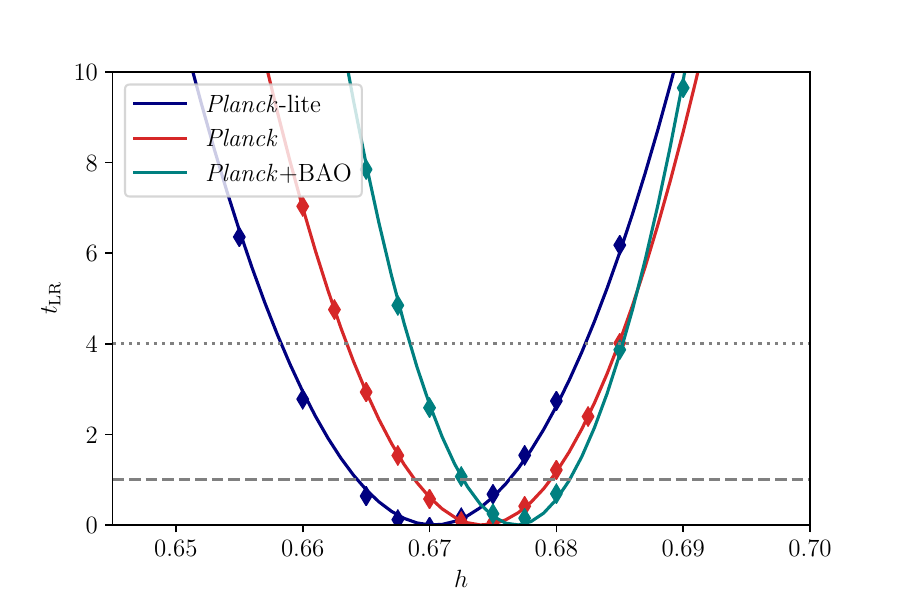}
    \caption{Profile likelihoods for the dimensionless Hubble parameter within $\Lambda$CDM, $h$, for \Planck-lite (blue), full \Planck\ (red), \Planck+BAO data (teal). The solid lines are parabolic fits to the data points. The $1\sigma$ ($2\sigma$) confidence intervals are obtained from the intersection of the profile likelihood with $t_\LR=1$ marked by the dashed grey line ($t_\LR=4$, dotted line).}
    \label{fig:PL_h}
\end{figure}
Constraints for all parameters of the $\Lambda$CDM model have been constructed in \cite{Planck:2013nga} for \Planck\ 2013 data, which found perfect agreement between Bayesian and frequentist constraints. Here, using updated \Planck\ data, we choose one parameter, the dimensionless Hubble parameter, $h$, which we constrain with the graphical profile likelihood method. The profile likelihoods for the three different data sets are shown in Fig.~\ref{fig:PL_h}. For a Gaussian distribution, $t_\LR$ corresponds to the usual $\Delta\chi^2$. Since our checks for the $\Lambda$CDM model in Sec.~\ref{sec:checks_LCDM} indicated that Wilks' theorem holds and $h$ does not have a physical boundary, we can use the simple graphical profile likelihood method to construct confidence intervals. For comparison, the posteriors of this model are shown in Fig.~\ref{fig:posterior_LCDM} in App.~\ref{sec:posteriors}. We summarize the frequentist and Bayesian constraints on $h$ in the first column of Tab.~\ref{tab:constraints}. We find good agreement between frequentist and Bayesian methods.

\subsection{Profiles in $M_\nu$}

Leaving the sum of neutrino masses, $M_\nu$ as a free parameter is a natural extension of the $\Lambda$CDM model. Curiously, \Planck\ data seems to favor ``negative'' $M_\nu$ (see \cite{Planck:2013nga} using profile likelihoods or more recently e.g.\ \cite{Green:2024xbb, Noriega:2024lzo, Naredo-Tuero:2024sgf}), making this model an interesting test case to study. Since the results of Sec.~\ref{sec:checks_Mnu} indicated that the distribution of $M_\nu$ is consistent with a Gaussian near a physical boundary, we use the boundary-corrected/Feldman-Cousins graphical construction (see Sec.~\ref{sec:Gaussian_boundaries}) to construct confidence intervals for $M_\nu$. 
\begin{figure}[h!]
    \centering
    \includegraphics[scale=0.64]{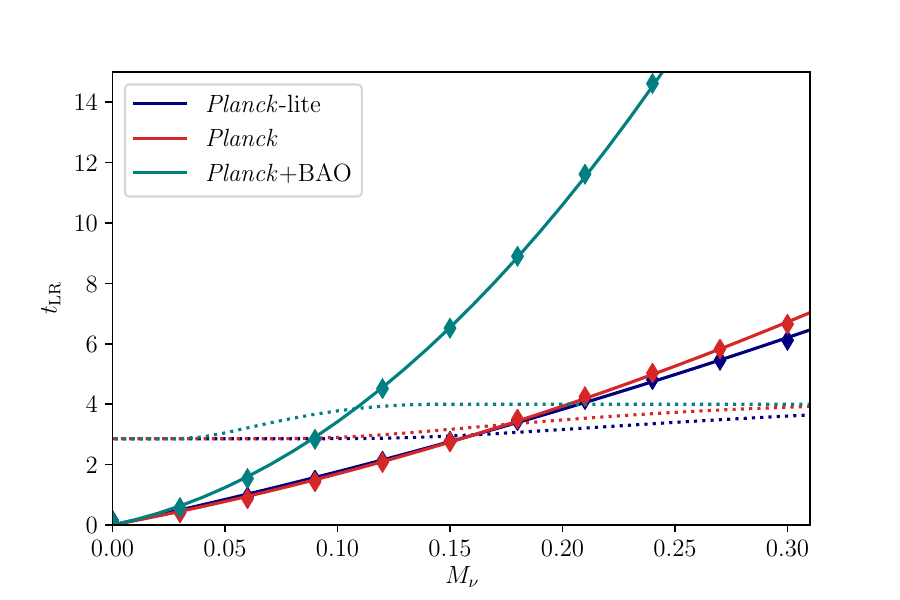}
    \includegraphics[scale=0.64]{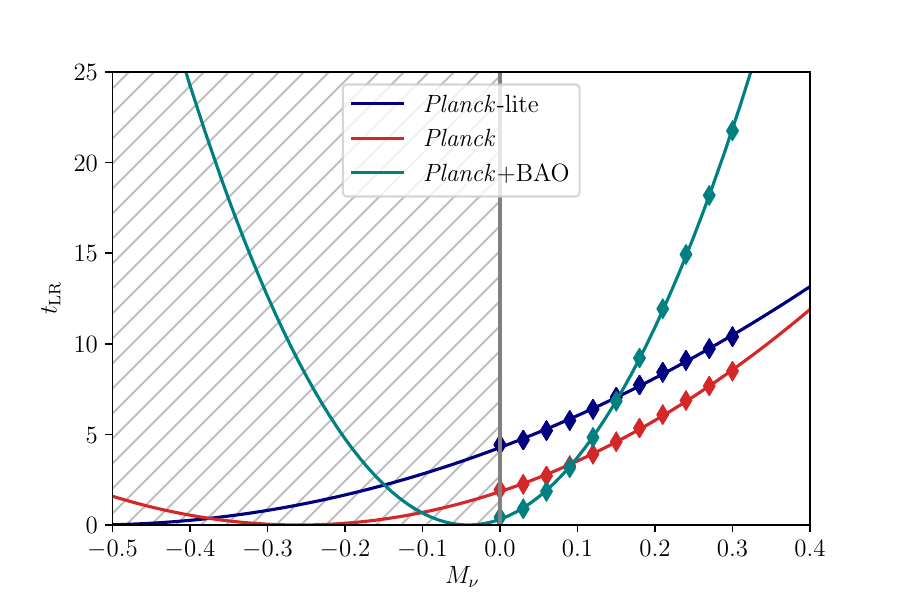}
    \caption{\textit{Top:} profile likelihoods for the sum of neutrino masses, $M_\nu$. Since $M_\nu$ is required to be non-negative and the best fit for all three data sets is $M_\nu = 0$, we use the boundary-corrected graphical construction. The corrected $2\sigma$ confidence intervals are given by the intersection of the dotted lines with the profile likelihoods of the respective colors. \textit{Bottom:} extrapolation of the parabolic fit to ``negative'' $M_\nu$. We vertically shift the profile likelihoods such that the minimum of the parabolic fit is in $t_\LR = 0$. This corroborates the apparent preference for negative $M_\nu$ found in previous work \cite{Planck:2013nga, Green:2024xbb, Naredo-Tuero:2024sgf}.}
    \label{fig:PL_Mnu}
\end{figure}

We show the profile likelihoods of $M_\nu$ in the top panel of Fig.~\ref{fig:PL_Mnu}. Since $M_\nu$ cannot be negative, there is a physical boundary in $M_\nu = 0$, and the global MLE of all three data sets is $\hat M_{\nu,\mathrm{phys}}=0$. The dotted lines show the boundaries of the $95\%$ acceptance region for the respective data set. These acceptance regions are obtained by adopting the likelihood test statistic, $t^\LR$ (Eq.~\ref{eq:tLR_boundary}), and assuming a Gaussian near a physical boundary. In practice, we obtain the acceptance regions for each data set by extrapolating the parabola fit to negative $M_\nu$ (bottom panel of Fig.~\ref{fig:PL_Mnu}) and determining $\sigma_{M_\nu}$ from the width of the extrapolated profile likelihood. Rescaling the confidence regions in Tab.~\ref{tab:corrected_graph_meth} in App.~\ref{sec:FC_table} by $\sigma_{M_\nu}$ gives the acceptance regions for the respective data sets (dotted lines).\footnote{This is equivalent to reading off $\hat M_\nu$ and $\sigma_{M_\nu}$ from the extrapolated profile likelihood and obtaining the boundary-corrected interval from \cite{Feldman:1997qc} (i.e.\ using $t=\hat M_\nu$ as a test statistic with ordering rule $t^\LR<t_0$). We checked that both approaches give the same result. In this case, the limits are far enough from the boundary that the naive graphical construction without any boundary correction gives the same result up to two significant digits.} 
%The limits sit close enough at the boundary, i.e.\ in the non-trivial part of the Neyman band, such that the boundary corrections become relevant.
For comparison, the posteriors of this model are shown in Fig.~\ref{fig:posterior_Mnu} in App.~\ref{sec:posteriors}.

We summarize the frequentist and Bayesian constraints on $M_\nu$ in the second column of Tab.~\ref{tab:constraints}. We find that the frequentist and Bayesian constraints are in broad agreement but the frequentist constraints are even tighter than the Bayesian constraints for \Planck\ and \Planck-lite data. Adding BAO data leads to a good agreement between the two approaches. 

Nevertheless, regardless of the statistical approach that is used, we obtain tight upper limits on $M_\nu$. To explore the observation that \Planck\ data seems to favor negative $M_\nu$, in the bottom panel of Fig.~\ref{fig:PL_Mnu}, we extrapolate the parabolic fit to negative neutrino masses. For \Planck\ and \Planck+BOSS data, we offset the parabola along the $y$ axis such that the minimum of the parabola lies in $t_\LR = 0$. 
We find that for \Planck\ and \Planck-lite data, the extrapolated minimum of the parabola lies far in the unphysical regime at $\sim\!-0.3\,$eV and $\sim\!-0.5\,$eV, respectively.
This leads to the tight upper limits $M_\nu < 0.18\,$eV and $M_\nu < 0.16\,$eV for \Planck\ and \Planck-lite data, respectively. Note that \Planck-lite data give a slightly tighter constraint since the standard deviation, i.e.\ the width of the parabola, is larger than for \Planck, which leads to an acceptance region for \Planck-lite (blue dotted line in the top panel of Fig.~\ref{fig:PL_Mnu}) that is slightly smaller than for \Planck\ (red dotted line). However, the results from the \Planck-lite data need to be taken with care since $A_{Planck}$ and $\tau_\reio$ have been fixed to their fiducial values as described in App.~\ref{sec:Pliklite}. 

This corroborates the apparent preference for negative neutrino masses in \Planck\ data, which was pointed out in e.g.\ \cite{Planck:2013nga, Green:2024xbb}. Once BAO data is added, the extrapolated minimum of the parabola shifts to $\sim\! -0.04\,$eV while the width of the parabola decreases. This leads to an even tighter limit of $M_\nu <0.011\,$eV.\footnote{This is in line with the tight upper limit from \Planck\ with DESI BAO data, which found $M_\nu<0.072$ \cite{DESI:2024mwx}, reminiscent of the apparent preference for negative $M_\nu$.} If this trend continues, these cosmological constraints pose a challenge to the inverted neutrino mass hierarchy.

While finalizing this work, \citep{Naredo-Tuero:2024sgf} appeared, which showed that, while \textit{Planck} 2018 \texttt{Plik} data \cite{Planck:2018vyg} (used in this work) appear to prefer negative $M_\nu$, this preference disappears with the new \Planck\ 2020 \texttt{HiLLiPoP} likelihood \cite{Tristram:2023haj} with both Bayesian and frequentist methods. Our results for \textit{Planck} 2018 \texttt{Plik} data are in good agreement with \citep{Naredo-Tuero:2024sgf}.

\subsection{Profiles in $w_0$}

Models with more complex dark energy than a cosmological constant have received increased attention recently due to the hint of time-evolving dark energy by the Dark Energy Spectroscopic Instrument (DESI, \cite{DESI:2024mwx}). Here, we assume a model with a constant equation of state $w_0$ as defined in Eq.~\eqref{eq:EOS_w0} as a free parameter. It is well known that \Planck\ data favors $w_0<-1$ \cite{Planck:2018vyg} and only adding BAO data shifts the constraints closer to $w_0 = -1$. This model was already studied with profile likelihoods with the Wilkinson Microwave Anisotropy Probe (WMAP, \cite{WMAP:2003elm, WMAP:2003xez, WMAP:2003ivt}) in \cite{Yeche:2005wn}. Here, we analyze this model with \Planck~\cite{Planck:2018vyg} and 6dF, SDSS, and BOSS data \cite{Beutler:2011hx,Ross:2014qpa,BOSS:2016wmc}.

\begin{figure}
    \centering
    \includegraphics[scale=0.64]{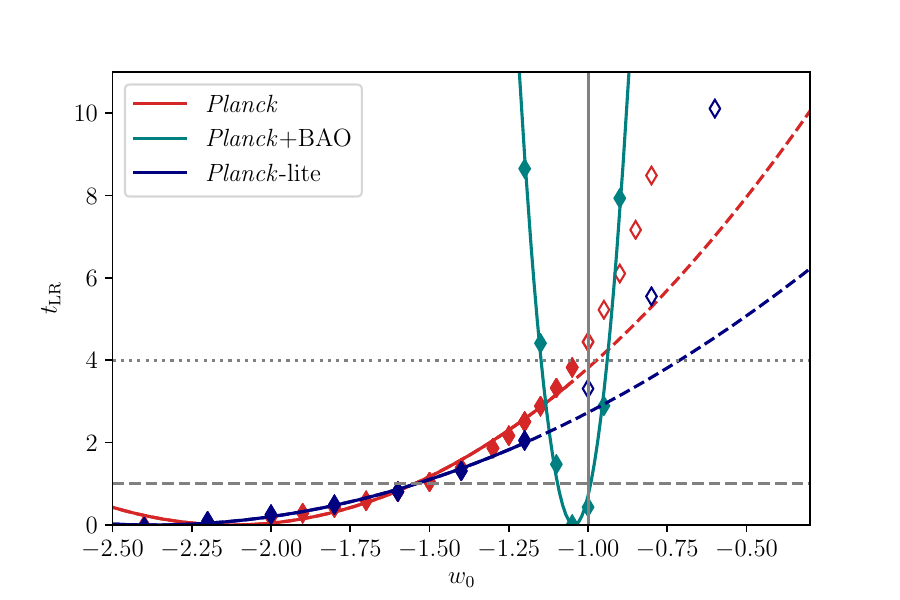}
    \caption{Profile likelihoods for the equation of state of DE, $w_0$, for different data sets. In Sec.~\ref{sec:checks_w0}, we found that the distribution of mocks is inconsistent with Wald's curve, which suggests that the graphical profile likelihood method does not give correct coverage. Regardless, we show $t_\LR=1$ (dashed line) and $t_\LR=4$ (dotted line), which give approximate $1\sigma$ and $2\sigma$ constraints.}
    \label{fig:PL_w0}
\end{figure}

We show the profile likelihoods of the $w_0$CDM model in Fig.~\ref{fig:PL_w0}. The deviation from Wald's curve that we found in Sec.~\ref{sec:checks_w0} suggests that the asymptotic assumption is not valid and the graphical profile likelihood construction does not yield correct coverage. Moreover, the profile likelihoods of $w_0$ in Fig.~\ref{fig:PL_w0} far away from the MLE are not well fit by a parabola and were excluded from the fit (open markers). Further, due to numerical difficulties in the Boltzmann solver \texttt{CLASS} \cite{Blas:2011rf}, very negative values of $w_0$ could not be explored and the interval construction relies on an extrapolation of the parabola to $w_0\lesssim -2$.

Therefore, to ensure coverage, the full Neyman correction would be necessary, which is beyond the scope of this paper. In the third column of Tab.~\ref{tab:constraints}, we quote the constraints obtained from the graphical profile likelihood method but acknowledge that they are only approximate. For comparison, the posteriors of this model are shown in Fig.~\ref{fig:posterior_w0} in App.~\ref{sec:posteriors}. The posteriors for \Planck\ and \Planck-lite are cut off by the prior boundaries at very negative $w_0$, so we quote upper limits in Tab.~\ref{tab:constraints}.

Our approximate constraints confirm that \Planck\ and \Planck-lite data favor very negative equations of state, $w_0<-2$. Only when adding BAO data, the degeneracy between $h$ and $w_0$ is broken (see Fig.~\ref{fig:posterior_w0}) and one receives a tight constraint, $w_0\approx -1.044 \pm 0.052$. For \Planck+BAO data, we find good agreement between our approximate frequentist and Bayesian constraints, as well as with previous constraints in the literature \citep{Planck:2018vyg, Semenaite:2022unt}. 

%%%%%%%%%%%%%%%%%%%%%%%%%%%%%%%%%%%%%%%%%%%%%%%%%%%%%%%%%%%%%%%%%%%%%%

\section{Conclusions}
\label{sec:conclusions}

Recently, there has been a growing interest in frequentist methods, particularly with the use of profile likelihoods, in various cosmological contexts. Despite its increased use, the graphical profile likelihood method relies on assumptions that are rarely checked. In this work, we reviewed profile likelihoods describing why, when, and how they can be used in cosmology, in particular focusing on testing the validity of the graphical method for different cases. This was illustrated for different models of interest in the context of \Planck\ CMB data.

We have reviewed the construction of frequentist confidence intervals. Although the Neyman construction can yield confidence intervals with correct frequentist coverage in any case, this construction is numerically expensive. When the asymptotic limit is reached and Wilks' theorem holds, in particular when the probability is Gaussian, the graphical profile likelihood construction can be used, where $68\%$ ($95\%$) confidence intervals are given by iso-likelihood contours obtained from the intersection of the profile likelihood with $\Delta \chi^2 = 1$ ($\Delta \chi^2 =4$). If the distribution is consistent with a Gaussian near a physical boundary, Wilks' theorem does not hold but confidence intervals with correct coverage can be obtained by the \textit{boundary-corrected} graphical construction or Feldman-Cousins construction. When neither of these cases apply, a full Neyman construction is in order. 

Reviewing these definitions is important since the standard graphical method is often used in the literature, even in situations where it can yield confidence intervals with incorrect coverage. Note that we do not view the frequentist framework as superior to the Bayesian one but rather complementary. Both frameworks can give misleading or wrong results if there is not enough data: while frequentist constraints might prefer ``fine-tuned'' models or coverage might not be fulfilled, Bayesian constraints can be subject to prior effects.\footnote{Note that coverage is guaranteed when using the full Neyman construction.} All these issues are remedied in the asymptotic limit of infinite data. However, if this limit is not reached -- as is often the case in science -- using both methods can give insights about the presence of these effects.

We illustrated these cases in cosmology in Sec.~\ref{sec:results}, testing the validity of the asymptotic assumption and Wilks' theorem for different models of interest, the $\Lambda$CDM, $\Lambda$CDM $+ M_\nu$, and $\omega_0$CDM. As expected, for the $\Lambda$CDM model for all the cosmological parameters, the distribution of the mock \textit{Planck}-lite spectra follows closely the predicted curves by Wilks and Wald for the fiducial cosmology. This indicates that the graphical profile likelihood construction gives correct coverage. 
For the $\Lambda$CDM $+ M_\nu$, the tests show that the distribution is compatible with a Gaussian near a physical boundary, and therefore, the boundary-corrected graphical construction should be used to obtain confidence interval with correct coverage. This model has been at the center of recent discussions in the literature regarding the preference for ``negative'' $M_\nu$ and sensitivity to the assumed prior. Therefore, obtaining meaningful confidence intervals for this model is crucial for the discussion and tests of this model.
The $\omega_0$CDM model is an important prototypical case where the distribution of \Planck-lite mocks does \textit{not} follow the curves by Wilks and Wald. In this case, where the distribution is not Gaussian and Wilks' theorem is violated, the (boundary-corrected) graphical construction will not guarantee correct coverage and a full Neyman construction is necessary to obtain meaningful confidence interval with correct coverage. 

We constructed frequentist confidence intervals for the Hubble parameter, $h$, within $\Lambda$CDM, for $M_\nu$, and for $w_0$ under \Planck-lite, \Planck, and \Planck+BAO data in Sec.~\ref{sec:profiles} and compared them to the Bayesian constraints: the intervals in $h$ show good agreement between both frameworks; for $M_\nu$, we found tighter upper limits in the frequentist framework, corroborating the apparent preference for negative $M_\nu$ in \Planck\ 2018 data; while for $w_0$ we expect that correct coverage might not be achieved, we regardless computed constraints with the graphical method, finding good agreement between frequentist and Bayesian constraints for \Planck+BAO. 

Since it is impractical to conduct these tests every time one uses the profile likelihood method, we provided some practical guidance in Sec.~\ref{sec:cookbook} on how to calculate frequentist confidence intervals and review their coverage. We made the \pinc\ code available, which can be used to compute the profile likelihoods as well as determine the (boundary-corrected) confidence intervals with the graphical construction. An extension of \pinc\ is left for future work.

There is a range of statistical methods at our hands in order to study the models and their behavior under data. The choice of method should be guided by the specific problem while keeping in mind the limitations and ranges of validity in the respective methods. All frameworks can be used to detect and understand unwanted or unknown effects in either framework. In this spirit, we believe that frequentist methods have a valuable place in cosmology as an additional tool to help extract maximal information from the remarkable cosmological data.

%%%%%%%%%%%%%%%%%%%%%%%%%%%%%%%%%%%%%%%%%%%%%%%%%%%%%%%%%%%%%%%%%%%%%%

\section*{Acknowledgements}
We are grateful to Graeme Addison for his help with the \Pliklite\ likelihood and to Sam Witte for his original suggestion to use simulated annealing minimization. We thank Graeme Addison, Eiichiro Komatsu, and Klaus Liegener for useful discussions and comments on the draft. LuH thanks Kyle Cranmer for teaching him the alternative Neyman construction visualizations. LaH would like to thank the Max Planck Institute for Astrophysics for the hospitality, where part of this work was conducted. Our analyzes were performed on the \texttt{freya} cluster maintained by the Max Planck Computing \& Data Facility. This research was supported by the Munich Institute for Astro-, Particle and BioPhysics (MIAPbP), which is funded by the Deutsche Forschungsgemeinschaft (DFG, German Research Foundation) under Germany's Excellence Strategy – EXC-2094 – 390783311.
Kavli IPMU is supported by the World Premier International Research Center Initiative (WPI), MEXT, Japan. EF thanks the support of the Serrapilheira Institute. LuH acknowledges support from the ORIGINS Cluster of Excellence funded by the Deutsche Forschungsgemeinschaft (DFG, German Research Foundation) under Germany's Excellence Strategy – EXC-2094 – 39078331.

%%%%%%%%%%%%%%%%%%%%%%%%%%%%%%%%%%%%%%%%%%%%%%%%%%%%%%%%%%%%%%%%%%%%%%

\appendix

\section{Expected standard deviation from Asimov data sets}
\label{sec:Asimov}

In the asymptotic regime, the values inferred from the mocks lie on the parabolic \textit{Wald's curve}, $t_{\onepoi}^\LR = (\hat{\onepoi}(x)-\onepoi_\true)^2/{\sigma_{\onepoi}^2}$ (Eq.~\ref{eq:Walds_relation}, \cite{Wald:1943}). The standard deviation $\sigma_\onepoi$ of this curve can be obtained from the so-called \textit{Asimov data set} \cite{Cowan:2010js}. This data set refers to a mock realization of the data, with all model parameters fixed to the ``true'' or fiducial parameters. 

Here, the Asimov data set corresponds to the CMB power spectra, $C_\ell$'s, computed with the Boltzmann solver \texttt{CLASS} \cite{Brinckmann:2018cvx} for the fiducial cosmology (without any statistical noise). The standard deviation $\sigma_\mathrm{Asimov}$ can be obtained as the $1\sigma$ confidence interval of the parameter of interest, $\onepoi$, under the Asimov data set, if appropriate e.g.\ by a graphical profile likelihood construction. 

\begin{figure*}
    \centering
    \includegraphics[scale=0.6]{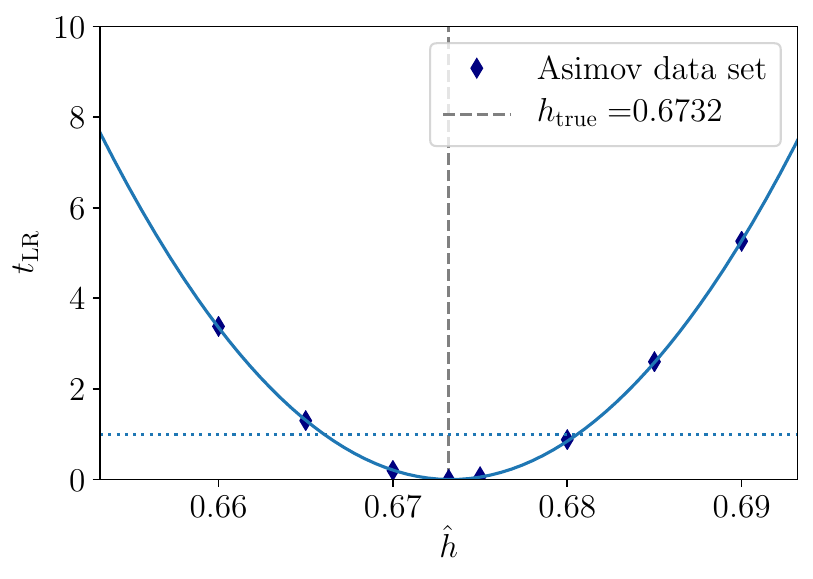}
    \includegraphics[scale=0.6]{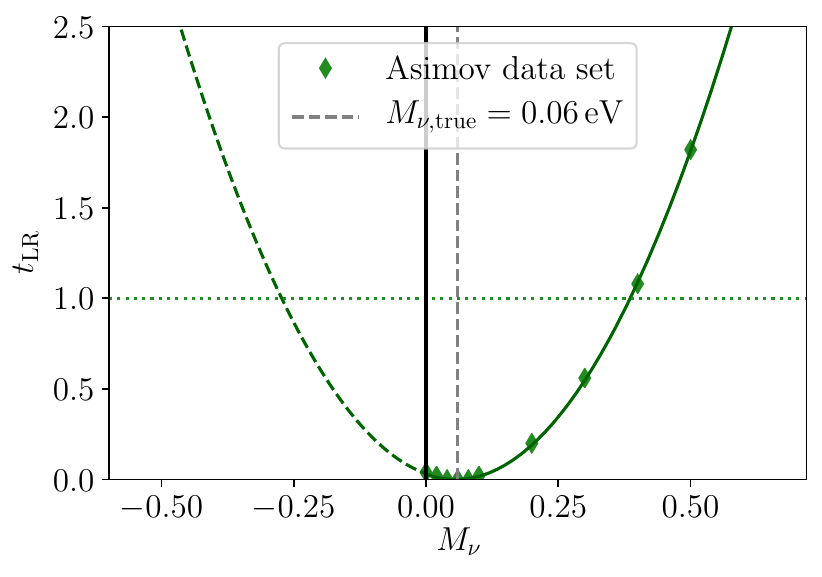}
    \includegraphics[scale=0.6]{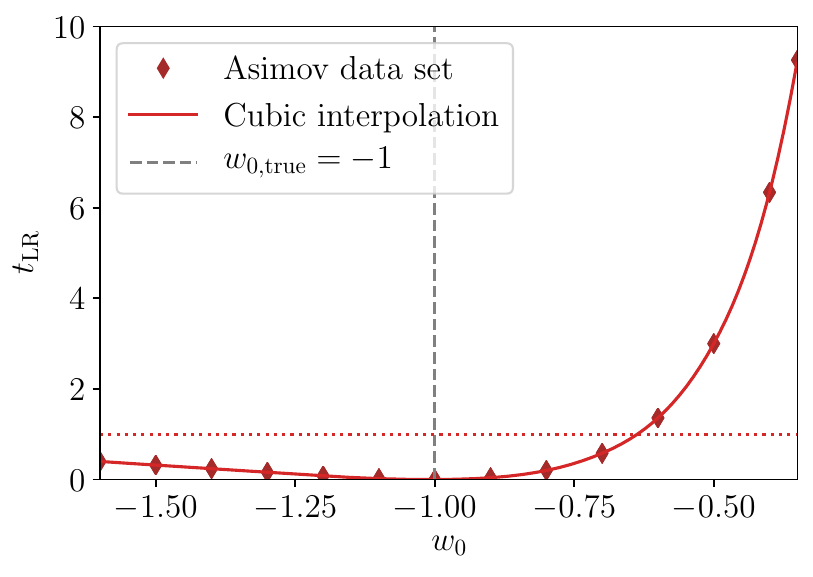}
    \caption{Profile likelihoods of the Asimov data sets, which are used to construct Wald's curve in Sec.~\ref{sec:results}. The profile likelihood of $h$ within $\Lambda$CDM corresponds to a parabola without any physical boundary. The profile likelihood of $M_\nu$  corresponds to a parabola with its minimum near a physical boundary in zero, and $w_0$ has a complicated asymmetric shape.}
    \label{fig:PLs_Asimov}
\end{figure*}

We show the profile likelihoods of the parameters of interest, here $h$, $M_\nu$, $w_0$ under the ``Asimov data set'' in Fig.~\ref{fig:PLs_Asimov}. From the profile likelihoods we obtain the $1\sigma$ standard deviation using the graphical profile likelihood method. The such obtained $\sigma_\mathrm{Asimov}^h$ and $\sigma_\mathrm{Asimov}^{M_\nu}$, are used to predict the expected distribution of mocks via Wald's relation in Sec.~\ref{sec:results}. 

The profile in $w_0$, however, deviates from a parabola due to the weak constraining power of \Planck(-lite) data for very negative $w_0$. This can be understood as being due to the degeneracy between $w_0$ and $h$, which is only broken when including BAO data, as can be seen in the posterior in Fig.~\ref{fig:posterior_w0}. Hence, we do not use the profile in $w_0$ to obtain $\sigma_\mathrm{Asimov}$. This deviation from a Gaussian likelihood in $w_0$ indicates that the graphical profile likelihood method will not yield correct coverage. 

\section{The \Pliklite\ binning scheme}
\label{sec:Pliklite}

In this appendix, we give details about the binning of the \Pliklite\ likelihood, which are necessary to generate mock \Pliklite\ data. The spectra $\hat{\vec{C}}$ and $\vec{C}(\vec \theta)$ (Eq.~\ref{eq:Plik_lite_like}) are summed into bins with width $\Delta \ell = 5$ for $30 \leq \ell \leq 99$ (14 bins), $\Delta \ell = 9$ for $100 \leq \ell \leq 1503$ (156 bins), $\Delta \ell = 17$ for $1504 \leq \ell \leq 2013$ (30 bins), and $\Delta \ell = 33$ for $2014 \leq \ell \leq 2508$ (15 bins)~\cite{Planck:2019nip}. This sums up to a total of 215 bins for TT and 199 for TE and EE. The bins are weighted according to (see Eq.~(22) in \cite{Planck:2019nip}):
\begin{equation}
    \label{eq:bins}
    C_b = \sum_{\ell = \ell_b^\mathrm{min}}^{\ell_b^\mathrm{max}} w_b^\ell C_\ell 
    \quad \mathrm{with} \quad
    w^\ell_b = \frac{\ell(\ell+1)}{\sum_{\ell = \ell_b^\mathrm{min}}^{\ell_b^\mathrm{max}} \ell (\ell +1)}.
\end{equation}

\section{Posteriors of \Pliklite\ and full \Planck\ data (+BAO)}
\label{sec:posteriors}

\begin{figure*}
    \centering
    \includegraphics[scale=0.6]{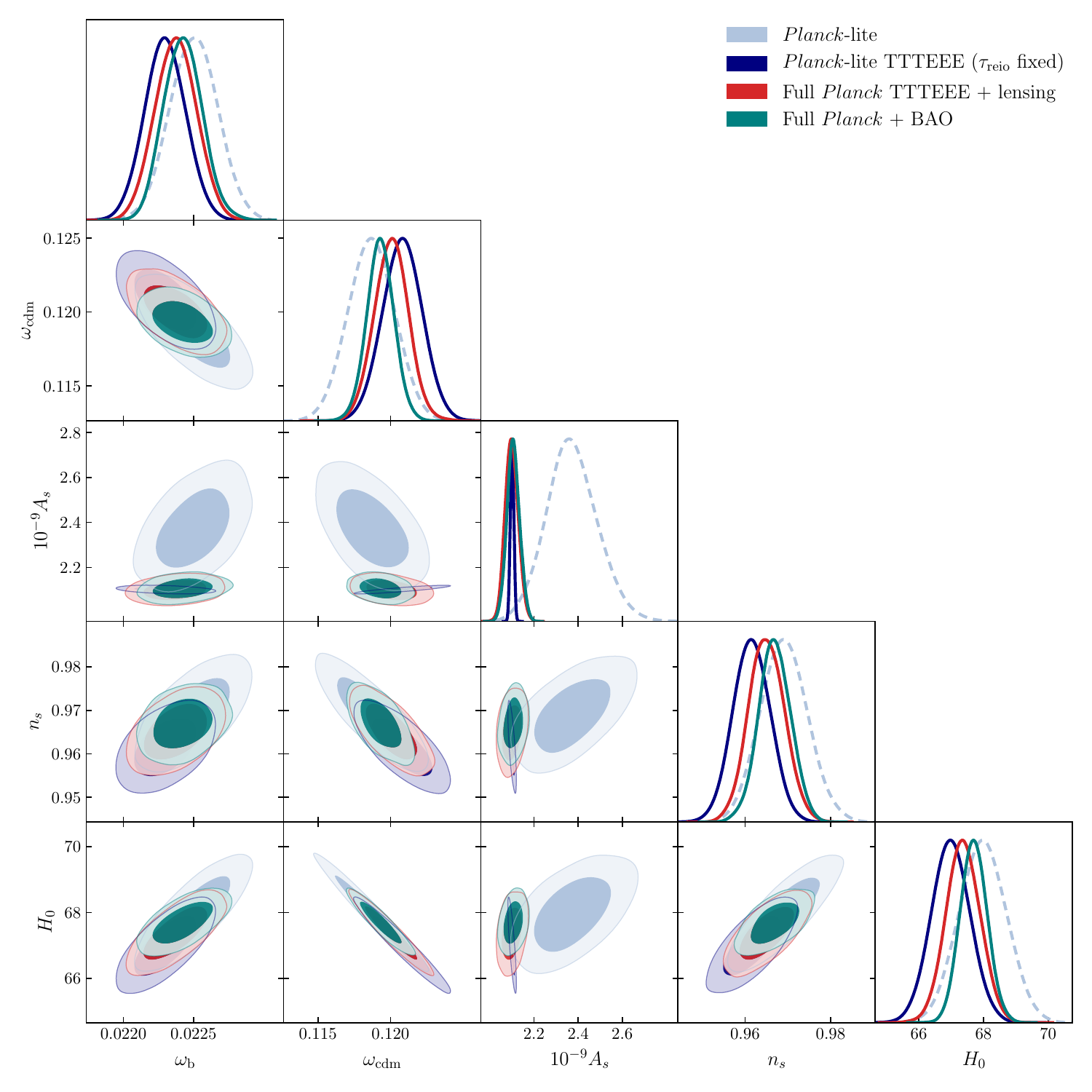}
    \caption{Posteriors of the $\Lambda$CDM parameters for four different data sets as indicated in the legend.}
    \label{fig:posterior_LCDM}
\end{figure*}

\begin{figure*}
    \centering
    \includegraphics[scale=0.6]{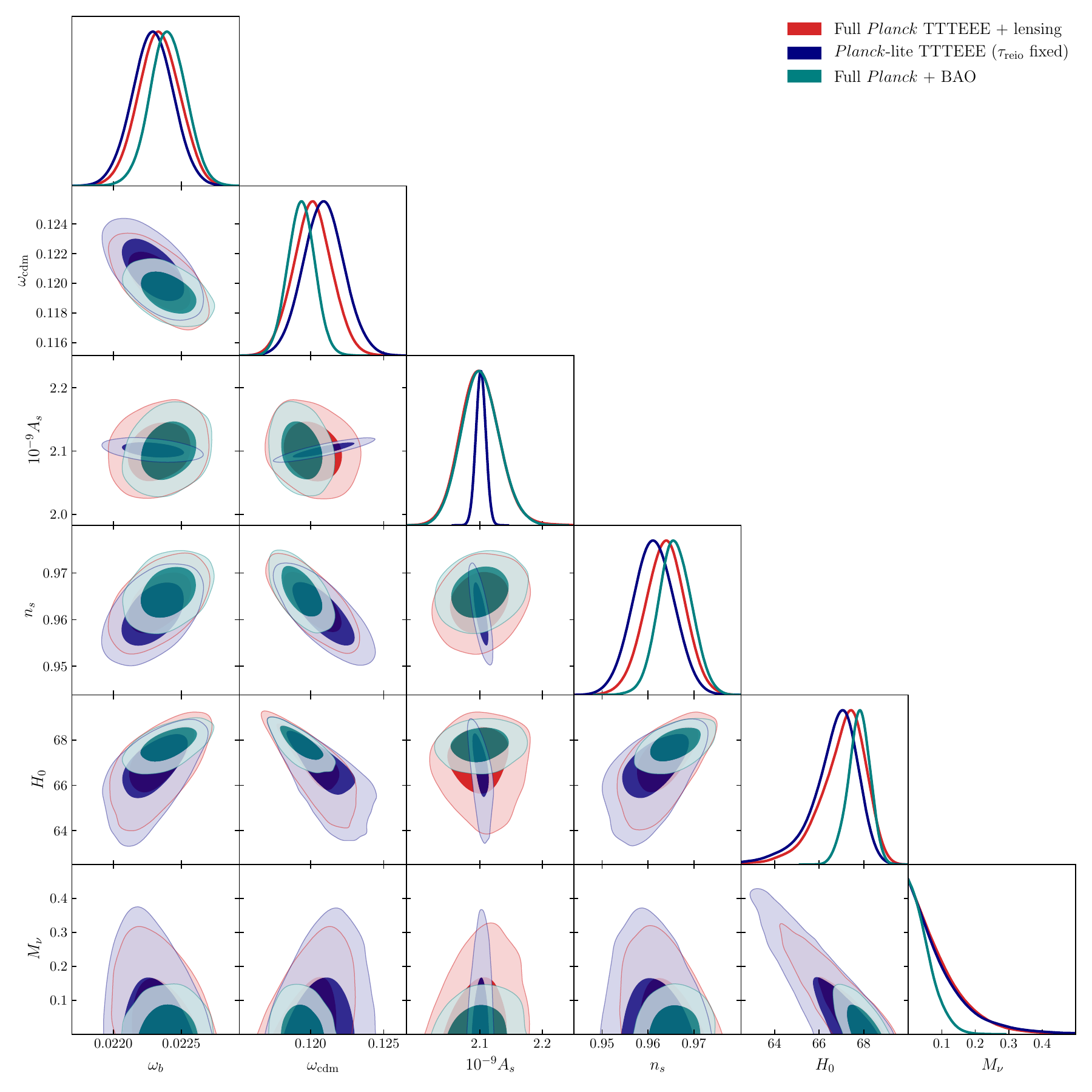}
    \caption{Posteriors of the $\Lambda$CDM+$M_\nu$ parameters for three different data sets as indicated in the legend.}
    \label{fig:posterior_Mnu}
\end{figure*}

\begin{figure*}
    \centering
    \includegraphics[scale=0.6]{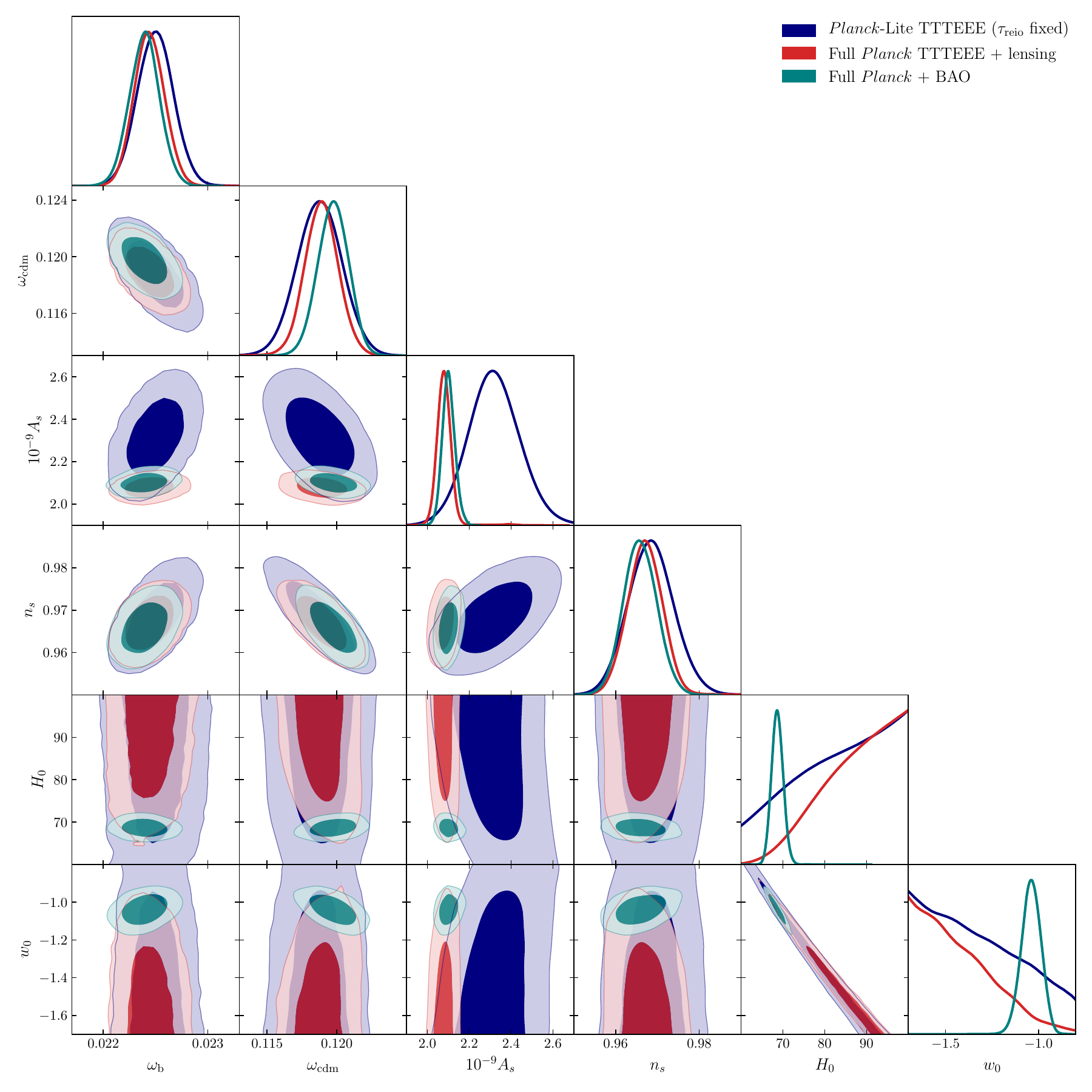}
    \caption{Posteriors of the $\Lambda$CDM+$w_0$ parameters for three different data sets as indicated in the legend.}
    \label{fig:posterior_w0}
\end{figure*}

We show the posteriors of the $\Lambda$CDM model in Fig.~\ref{fig:posterior_LCDM}, of the $\Lambda$CDM+$M_\nu$ model in Fig.~\ref{fig:posterior_Mnu}, and of the $\Lambda$CDM+$w_0$ model in Fig.~\ref{fig:posterior_w0}. We consider the same data set combinations as for the profile likelihoods in Sec.~\ref{sec:profiles}: the pre-marginalised \Pliklite\ TTTEEE data (with $\tau_\mathrm{reio}$ fixed), the full \Planck\ TTTEEE and lensing data, and the full \Planck\ data combined with BAO data from 6dFGS, SDSS and BOSS (see Sec.~\ref{sec:data_and_mocks} for details). For the $\Lambda$CDM model (Fig.~\ref{fig:posterior_LCDM}), we additionally show the \Pliklite\ TTTEEE likelihood with $\tau_\mathrm{reio}$ as a free parameter. We require the Gelman-Rubin criterion $R-1<0.05$.

\section{Likelihood ratio histograms in $\Lambda$CDM for all cosmological parameters}
\label{sec:LR_hist_LCDM_fixed_NP}

\begin{figure*}[htp]
    \centering
    \subfigure{\includegraphics[scale=0.5]{LR_hist_fixed_NP_h.pdf}}\quad
    \subfigure{\includegraphics[scale=0.5]{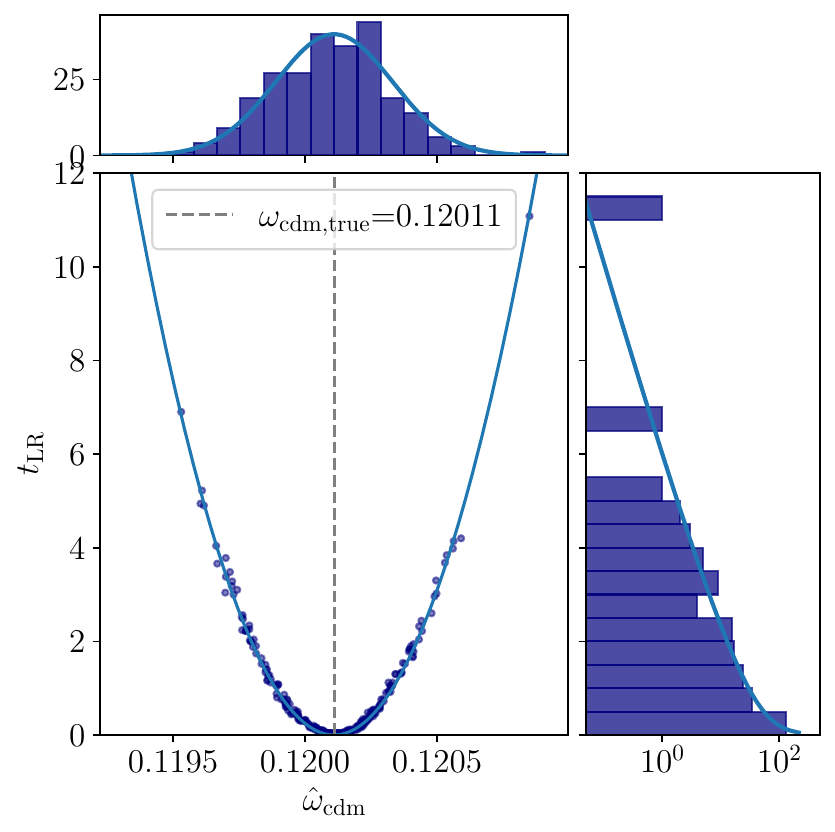}}
    \subfigure{\includegraphics[scale=0.5]{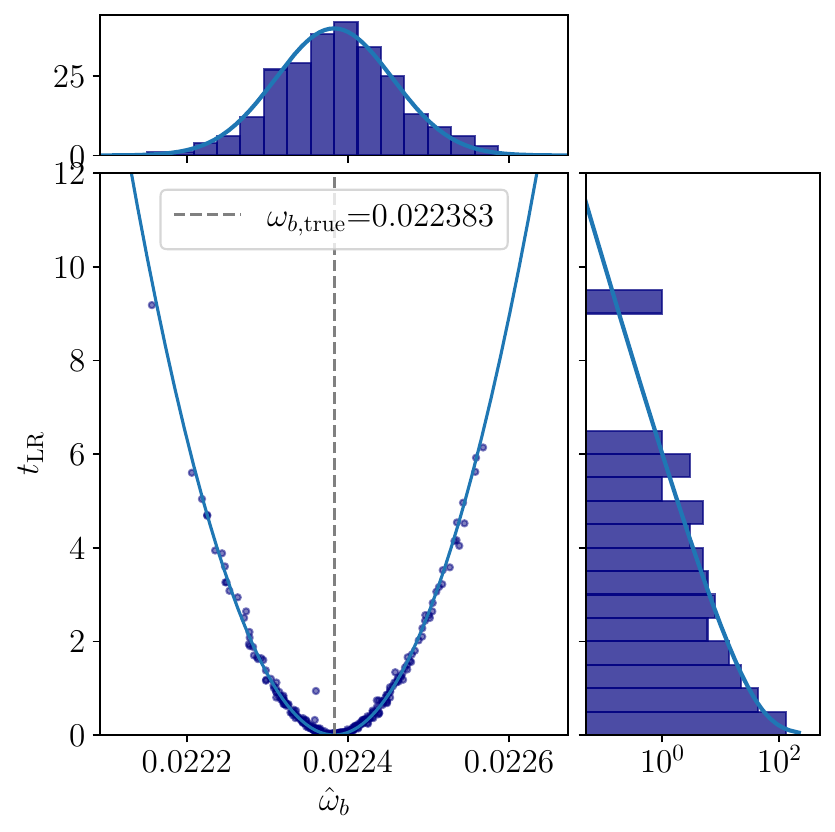}}\quad
    \subfigure{\includegraphics[scale=0.5]{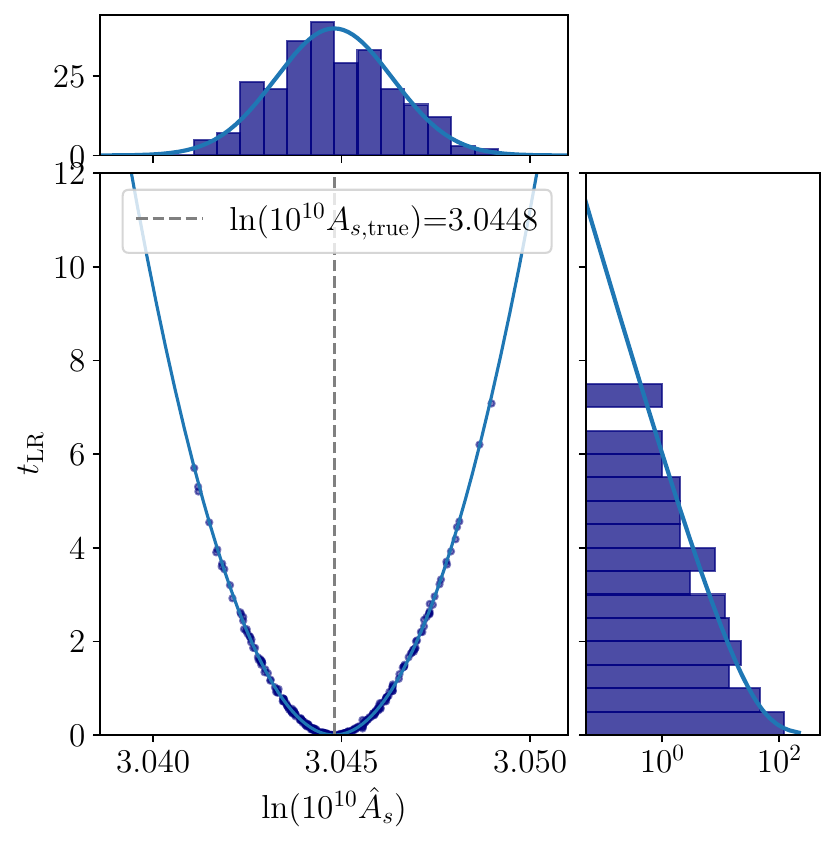}}
    \subfigure{\includegraphics[scale=0.5]{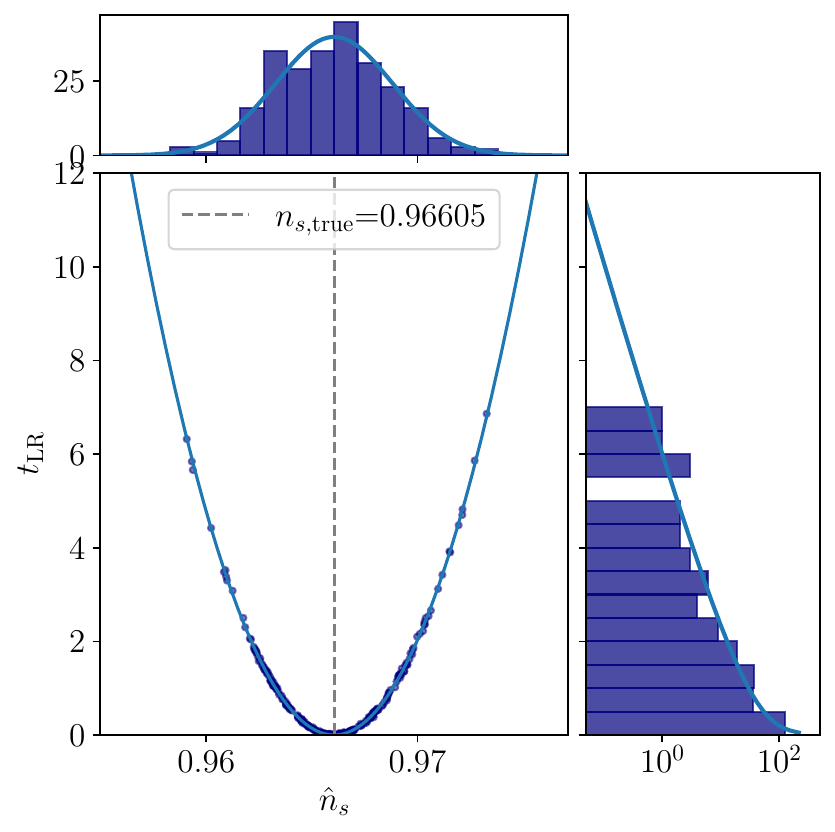}}\quad
    \subfigure{\includegraphics[scale=0.5]{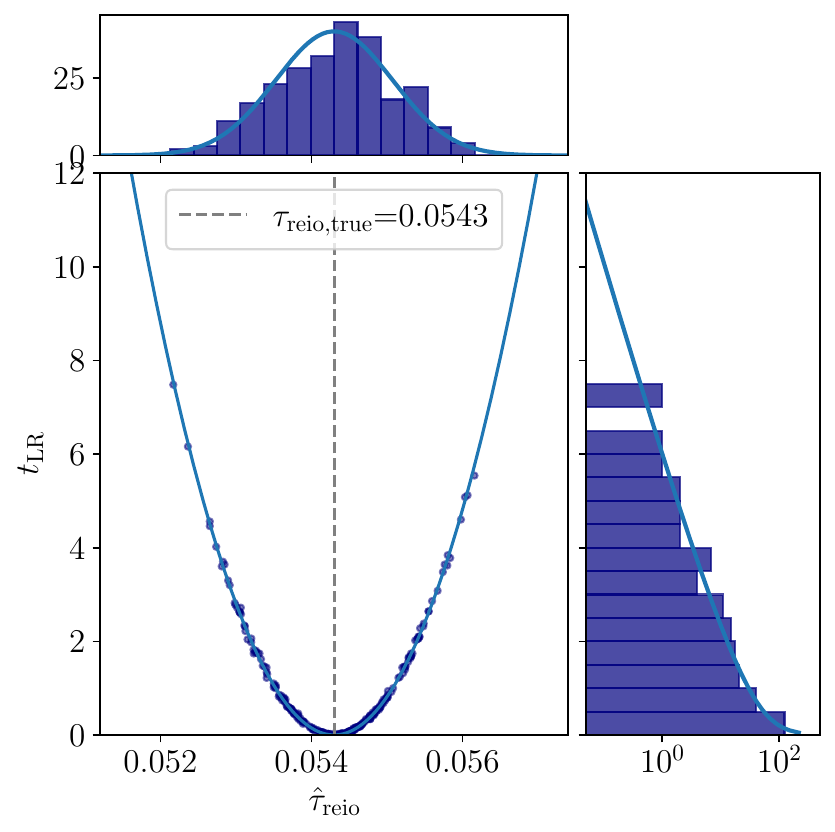}}
    \caption{Each of the six panels shows the LR test statistic $t_\mathrm{LR}$, Eq.~\eqref{eq:LR_mocks}, for one of the six \LCDM\ parameters, while holding all other parameters fixed. The markers show $t^\LR$ as a function of the MLE of the respective parameter obtained from 250 mock realizations of the \textit{Planck}-lite data, which follow closely ``Wald's curve'' as indicated by the solid blue line, where the vertical grey dashed line shows the true value of the parameter. The histogram along the $x$ axis shows the number of mocks in bins of the \LCDM-parameter of interest, which is consistent with a Gaussian distribution as indicated by the solid blue line. The histograms along the $y$ axis show the number of mocks in bins of $t^\LR$, which is consistent with a $\chi^2$ distribution as indicated by the solid blue line.}
    \label{fig:LR_hist_LCDM_fixed_NP}
\end{figure*}

The six panels in Fig.~\ref{fig:LR_hist_LCDM_fixed_NP} show the LR test statistic $t^\LR$ in Eq.~\eqref{eq:LR_mocks}, where $\onepoi$ is one of the six \LCDM\ parameters as a function of the \textit{inferred} MLE of the parameter of interest, $\hat \onepoi$, while the remaining cosmological parameters are kept fixed. The first panel in Fig.~\ref{fig:LR_hist_LCDM_fixed_NP} is the same as the top panel in Fig.~\ref{fig:LR_hist_LCDM}. 
 
\section{Table for boundary-corrected graphical profile likelihood method}
\label{sec:FC_table}

For a Gaussian near a physical boundary, the boundary-corrected graphical profile likelihood method can be used, which we review in Sec.~\ref{sec:Gaussian_boundaries}. For convenience, we add here the corrected $1\sigma$ and $2\sigma$ confidence intervals for a Gaussian near a physical boundary at zero in Tab.~\ref{tab:corrected_graph_meth}, which we used in Fig.~\ref{fig:PL_Mnu} and Sec.~\ref{sec:profiles}. The mean $\mu$ of the Gaussian is quoted in units of the standard deviation $\sigma$. The corrected $68\%$ ($95\%$) confidence interval is given at the intersection of the profile likelihood with the interpolation of the respective column. For $\mu$ far away from the physical boundary at zero, the familiar cutoff at $t_\LR=1$ ($t_\LR=4$) is recovered. 
\begin{table}[t!]
    \centering
    \begin{tabular}{c|c|c}
    $\mu$   &$68\%$ C.L.     &$95\%$ C.L.\\
    \hline
    0.00    &0.23     &2.86 \\ 
    0.05    &0.25     &2.86 \\ 
    0.10    &0.34     &2.86 \\ 
    0.15    &0.44     &2.86 \\ 
    0.20    &0.52     &2.86 \\ 
    0.25    &0.59     &2.86 \\ 
    0.30    &0.65     &2.86 \\ 
    0.35    &0.71     &2.86 \\ 
    0.40    &0.76     &2.86 \\ 
    0.45    &0.80     &2.87 \\ 
    0.50    &0.84     &2.89 \\ 
    0.55    &0.87     &2.92 \\ 
    0.60    &0.90     &2.96 \\ 
    0.65    &0.93     &3.01 \\ 
    0.70    &0.95     &3.07 \\ 
    0.75    &0.96     &3.13 \\ 
    0.80    &0.98     &3.19 \\ 
    0.85    &0.99     &3.25 \\ 
    0.90    &0.99     &3.31 \\ 
    0.95    &1.00     &3.37 \\ 
    1.00    &1.00     &3.43 \\ 
    1.05    &1.00     &3.48 \\ 
    1.10    &1.00     &3.54 \\ 
    1.15    &1.00     &3.59 \\ 
    1.20    &1.00     &3.64 \\ 
    1.25    &1.00     &3.68 \\ 
    1.30    &1.00     &3.72 \\ 
    1.35    &1.00     &3.76 \\ 
    1.40    &1.00     &3.80 \\ 
    1.45    &1.00     &3.83 \\ 
    1.50    &1.00     &3.86 \\ 
    1.55    &1.00     &3.89 \\ 
    1.60    &1.00     &3.91 \\ 
    1.65    &1.00     &3.93 \\ 
    1.70    &1.00     &3.95 \\ 
    1.75    &1.00     &3.97 \\ 
    1.80    &1.00     &3.98 \\ 
    1.85    &1.00     &3.99 \\ 
    1.90    &1.00     &3.99 \\ 
    1.95    &1.00     &4.00 \\ 
    2.00    &1.00     &4.00 \\ 
    \end{tabular}
    \caption{Tabulated values of the cutoff in $t_\LR = \Delta\chi^2$ of a Gaussian near a physical boundary for $68\%$ and $95\%$ C.L., respectively, as a function of the model parameter $\mu$ in units of the standard deviation $\sigma$. This is used to define the acceptance regions in Fig.~\ref{fig:PL_Mnu}.}
    \label{tab:corrected_graph_meth}
\end{table}

%%%%%%%%%%%%%%%%%%%%%%%%%%%%%%%%%%%

\bibliography{main}
\bibliographystyle{apsrev4-1}

\end{document}